\documentclass[a4paper,12pt]{article}

\usepackage{amssymb,latexsym}

\makeatletter \@addtoreset{equation}{section} \makeatother

\begin{document}

\begin{titlepage}

\thispagestyle{empty}

\begin{flushright}
\hfill{HU-EP-03/75} \\
\hfill{hep-th/0311146}
\end{flushright}

\vspace{35pt}

\begin{center}{ \LARGE{\bf
${\cal N} =1$ geometries for M--theory \\[4mm]
and type IIA strings  
with fluxes}}

\vspace{60pt}

{\bf  Gianguido Dall'Agata \  and \ Nikolaos Prezas}

\vspace{15pt}

{\it  Humboldt Universit\"at zu Berlin,
Institut f\"ur Physik,\\
Newtonstrasse 15, D-12489 Berlin, Germany}\\[1mm] {E-mail:
dallagat, prezas@physik.hu-berlin.de}

\vspace{50pt}

{ABSTRACT}

\end{center}

\vspace{20pt}

We derive a set of necessary and sufficient conditions for obtaining 
${\cal N}= 1$  backgrounds of M--theory and type IIA 
strings in the presence of fluxes. 
Our metrics are warped products of four--dimensional Minkowski
space--time with a curved internal manifold.
We classify the different solutions for irreducible internal manifolds
as well as for manifolds with $S^1$ isometries by employing the
formalism of group structures and intrinsic torsion.
We provide examples within these various classes along with general techniques 
for their construction.
In particular, we generalize the Hitchin flow equations so that one 
can explicitly build irreducible 7--manifolds with 4--form flux.
We also show how several of the examples found in the literature fit in our
framework and suggest possible generalizations.

\end{titlepage}

\newpage

\baselineskip 6 mm

\section{Introduction}

The need to connect ordinary four--dimensional physics
with string theory or M--theory motivates the study of all types of
solutions which can be described as a (possibly warped) product of
four--dimensional Minkowski space--time with an internal 6-- or
7--manifold.
Although one would finally need to completely break supersymmetry, 
retaining some control on the effective theory suggests 
to look for solutions leading to ${\cal N} = 1$ 
supersymmetry in four--dimensions.

The simplest such scenario consists in a setup where all fields but the metric 
are vanishing.
There, it is known that the resulting internal space must 
fall in Berger's classification of special holonomy manifolds.
Being more specific, in M--theory one uses 
7--manifolds of $G_2$ holonomy, whereas in the context of the 
heterotic string theories one needs Calabi--Yau three--folds, whose 
holonomy is $SU(3)$. Although 
these solutions may be interesting phenomenologically, one can also
consider the more general case where other fields besides the
metric acquire non--vanishing expectation values.
This is actually very natural in string theory, where it is known that
D--branes couple to the various tensor fields appearing in the
theory.
Therefore, an obvious extension consists in the analysis of string-- and 
M--theory vacua in the presence of non--trivial fluxes, i.e. 
non--vanishing expectation values for the tensor fields.
Moreover, in recent times, the inclusion of fluxes has provided  
new insight in addressing the moduli problem 
(see for instance \cite{Frey:2003tf} and references therein) and in  
constructing potentials leading to $dS$ vacua.

Although the concept of holonomy is no longer a very useful tool
for classifying these types of solutions, it can be shown that an analogous  
r\^{o}le is now played by group structures.
It was already noted in the `80's that the requirement of 
supersymmetric solutions implies the existence of
tensor structures given by bilinears in the supersymmetry parameters 
\cite{uno}.
One outstanding example is given by the K\"ahler structure
$
{J_m}^n = {\rm i } \,\eta^\dagger {\gamma_m}^n \eta\,
$
of Calabi--Yau manifolds
where $\eta$ is the supersymmetry parameter. 
This fact was reconsidered in \cite{Gauntlett:2001ur}, where
a more precise connection with the group structure of 
the solution was remarked.
Indeed, if such tensors are globally defined, they imply a reduction of 
the structure group of the tangent bundle. 
As a consequence, the supersymmetry requirement on the solutions can be reinterpreted 
as a restriction on the possible group structures.
As a final outcome, it is quite important to
know and to classify group structures not only as a way to extract general 
information on the solutions but also in order to find techniques for their 
construction.

So far, the main effort in studying flux compactifications  
has been devoted to type IIB 
\cite{collect1,Kachru:2002sk,Tripathy:2002qw}
and heterotic theories 
\cite{collect2,Cardoso:2002hd,Gauntlett:2003cy}.
In these cases, it was shown that the internal manifold is no longer 
Calabi--Yau, but retains the property of being a complex manifold.
On the other hand, not so much has been done for the type IIA 
theory, despite its prominent r\^ole in connection with  
intersecting brane--world scenarios (see 
\cite{Ibanez:2001dj,Blumenhagen:2002vp,Cvetic} and references therein).
However, type IIA solutions can be obtained from circle reductions of
M--theory and as such they appear indirectly in the analysis of all
possible supersymmetric solutions of M--theory 
\cite{Gauntlett:2002fz,Gauntlett:2003new}.
Moreover, always in the context of M--theory, a classification of the 
solutions we seek in terms of $G_2$ structures 
has been given in \cite{Behrndt:2003uq,Kaste:2003zd}\footnote{Solutions with a 
3--dimensional Minkowski space--time were analyzed in 
\cite{collect3}
.} and \cite{Kaste:2002xs,Kaste:2003dh,Behrndt:2003ih} 
analyzed the $SU(3)$ structures of 
six--dimensional manifolds that can be used for type IIA compactifications
with two--form flux.

Despite these very general classifications, no explicit examples nor guideline 
for their construction were given in these papers.
For this reason, we would like to take a more concrete approach
and analyze once more M--theory in the presence of fluxes. In particular,
by performing a systematic analysis 
of the various classes of group structures allowed by  
supersymmetry, we will finally be able to produce explicit examples.
In doing so, we are not only going to discuss the purely 
eleven--dimensional backgrounds, but also the various possible IIA 
reductions.

An important mathematical result in this vein 
is that a spin seven--manifold  admits always an $SU(2)$ structure
\cite{Friedrich:1995Dp}.
Hence, it would seem natural at first sight to classify the 
supersymmetric solutions of M--theory according to their  $SU(2)$ structure.
However, this turns out to give a very complicated rewriting of the susy 
conditions that makes further analysis quite cumbersome.
For this reason we choose to consider an intermediate setup and analyze 
$SU(3)$ structures. 
We will see that this analysis is fine enough to capture the main
properties of the supersymmetric solutions and contains the guidelines
for the construction of explicit examples.

By using this strategy, 
we find the necessary and sufficient conditions 
for obtaining supersymmetric solutions, in terms of restrictions 
on the 4--form flux and the intrinsic torsion of the internal manifold.
We then use this result to study the two main classes 
of solutions: irreducible 7--manifolds and manifolds with an 
$S^1$ isometry.

Regarding the first class, we find two interesting results.
First, it is possible to construct 
a generalization of the Hitchin flow equations \cite{Hitchin}.
In the same way as the Hitchin construction  
yields $G_2$--holonomy manifolds fibering half--flat 6--manifolds over 
an interval $I\subseteq {\mathbb R}$ \cite{chiossi}, 
we construct 7--manifolds with the appropriate $SU(3)$ 
structure starting from special--hermitian manifolds. 
Second, by analyzing a more general setup in which 
the einbein over $I$ depends on the $M_6$ 
coordinates, we are able to 
recover the Fayyazuddin--Smith solution
\cite{Fayyazuddin:1999zu,Brinne:2000nf,Smith:2002wn,Husain:2003df}.

Subsequently, we consider 7--manifolds with an $S^1$ isometry, where we can 
further distinguish two classes of solutions.
These arise because of the presence of a vector $v$ 
in the definition of an $SU(3)$--structure in 7 
dimensions.
Then, one can distinguish between reductions to type IIA where $v$ is
respectively proportional or orthogonal to the Killing vector
describing the $S^1$ isometry along which we reduce M--theory.
In the first instance we find that the type IIA theory 
can be described in terms of $SU(3)$ classes in six dimensions.
More importantly, it contains only the NS 3--form flux and therefore it
gives rise to the known results of \cite{Cardoso:2002hd,Gauntlett:2003cy}, 
where the common sector of type I, type II and heterotic string theories was analyzed.
Conversely, when the isometry is ``orthogonal'' to $v$, we can further 
refine the analysis according to the type of fluxes one obtains in 
10 dimensions.
An interesting result is that without 4--form flux and appropriately
chosen 2--form and 3--form fluxes, one can use conformal Calabi--Yau
manifolds to compactify type IIA string theory to four dimensions.
As a final application we provide a technique to build vacua of type 
IIA with all fluxes and dilaton turned on starting from ${\mathbb 
T}^2$ fibrations over $K3$ manifolds.

The plan of the paper is the following.
After this introduction, in section 2 we review $SU(3)$ and $SU(2)$ 
structures for 6-- and 7--dimensional manifolds and we describe how 
one can classify the various possibilities in terms of the irreducible 
modules of the intrinsic torsion.
In section 3 we recall the conditions to obtain supersymmetric 
solutions of M--theory with non--vanishing 4--form flux and express 
these in terms of $SU(3)$ structures, providing a set of necessary and 
sufficient conditions the flux and the internal manifold should obey.
The construction of explicit examples and the description of general 
techniques to obtain them starts in section 4, where we analyze 
irreducible 7--manifolds.
We show how to obtain them as fibration of 6--manifolds on real 
intervals giving generalizations to the Hitchin construction of 
$G_2$--holonomy manifolds and recovering the Fayyazuddin--Smith 
solution of M5--branes wrapped on holomorphic 2-cycles of the internal
manifold.
Finally, in section 5 we discuss the type IIA reduction for  
7--manifolds admitting isometries, making contact with known results 
and discussing new possibilities arising from turning on all possible 
fluxes.

\medskip

{\bf Note added:} While this paper was under completion we were 
informed of the work by Behrndt and Jeschek \cite{ClausKlaus} 
which has some overlap with our section \ref{irredI} and \ref{redv}, and discusses also 
the superpotentials for M--theory with 4--form flux.
A refined discussion of type II theories with NS--fluxes and the 
relation with mirror symmetry appeared in \cite{last}.

\section{Group structures and torsion classes}

The existence of a $G$--structure on a $d$--dimensional Riemannian
manifold implies that the structure group of the frame bundle can be
reduced to $G\subset O(d)$ (if the manifold is spin then $G \subset 
Spin(d)$).  
An alternative and sometimes more convenient way to define
$G$--structures is via $G$--invariant tensors (spinors).
A non--vanishing, globally defined tensor (spinor) $\eta$ is $G$--invariant if it is invariant
under $G$--rotations of the orthonormal frame. 
Since $\eta$ is globally defined, by considering the set of frames for which $\eta$
takes the same form, one can see that the structure group of the
frame bundle reduces to $G$ or a subgroup thereof. 
Thus the existence of $\eta$ implies a $G$--structure. 

Typically, the converse is also true.
Tensors of a given type, relative to an orthonormal frame, form a
vector space, or module, for a given representation of $O(d)$.
If the structure group of the frame bundle is reduced to $G\subset
O(d)$, this module can be decomposed into irreducible modules of $G$.
If there are tensors admitting invariant components under 
$G$, the corresponding vector bundle must be trivial, and
thus it will admit a globally defined non--vanishing section $\eta$.   

The existence of a $G$--structure does not a priori put any constraints
on the possible holonomy groups. 
In particular, the failure of the holonomy of the Levi-Civita
connection to reduce to $G \subset GL(n)$ is measured by the intrinsic
torsion and this latter can be used to describe 
the $G$--structure.
Given some $G$--invariant form $\eta$ defining a $G$--structure, the
derivative of $\eta$ with respect to the Levi--Civita connection,
$\nabla\eta$, can be decomposed into $G$--modules.
The different types of $G$--structures are then specified by which of
these modules are present, if any.
One first uses the fact that there is no obstruction to 
find some connection $\nabla^{(T)}$ so that $\nabla^{(T)}\eta=0$ 
\cite{Joyce}. 
Then $\nabla^{(T)}-\nabla$ is a tensor which has values
in $\Lambda^1\otimes \Lambda^2$. 
Since $\Lambda^2\cong so(d)=g\oplus g^\perp$
where $g^\perp$ is the orthogonal complement of the Lie algebra $g$ in 
$so(d)$, and $\eta$ is invariant with respect to $g$,
we conclude that $\nabla\eta =(\nabla-\nabla^{(T)})\eta$ can be identified with
an element $\tau$ of $\Lambda^1\otimes g^\perp$.
Furthermore, this element is a function only of the particular
$G$--structure, independent of the choice of $\nabla^{(T)}$ and 
it is in one-to-one correspondence with the intrinsic torsion.
Explicitly, for a $p$--form $\eta$
\begin{equation}
\nabla_m \eta_{n_1\dots n_p}=
-p \;\tau_{m\,[n_1}{}^q\,\eta_{|q|n_2\dots n_p]}\,,
\end{equation}
where $\tau \in\Lambda^1\otimes g^\perp$,  $m$ is the one--form 
index and $n$, $q$ label the
two--form $g^\perp\subset\Lambda^2$. 


The search for supersymmetric solutions of string and supergravity 
theories demands the existence of spinors which annihilate all the 
supersymmetry transformations.
In geometrical terms, such spinors are parallel with 
respect to a generalized connection which include the Levi--Civita 
connection and the fluxes contributions:
\begin{equation}
\nabla^{(T')}\eta = 0\,.
\label{susyreq}
\end{equation}
This gives us the possibility of understanding whether a certain 
solution preserves supersymmetry or not by analyzing its group 
structure in terms of the intrinsic torsion.
Indeed one needs its group--structure to be contained in those allowed by (\ref{susyreq})
\begin{equation}
\nabla^{(T)}\subseteq\nabla^{(T')}\,.
\label{ssycon}
\end{equation}
It is therefore very important to express supersymmetry conditions 
as constraints on the intrinsic torsion and at the same time to  
classify the possible group--structures of the candidate 
solutions in terms of the irreducible components of the same intrinsic torsion.
This is still not enough to satisfy the equations of motion, 
unless one requires maximal supersymmetry.
As we will see later, only in certain favorable cases one can 
translate the extra conditions coming from such a requirement in terms of 
torsion classes. We will always try to achieve 
this, so that specifying the group structure is everything 
one needs in order to completely satisfy all the  conditions.

So far we assumed that the supersymmetry parameter is a  
spinor and that in order 
to fulfill the supersymmetry conditions one has to use spin manifolds.
Actually, in certain cases a weaker requirement can guarantee the
existence of (locally) supersymmetric solutions. 
There are cases where a $Spin_c$--structure is enough.
If this happens\footnote{We thank D. 
Martelli for explaining this to us.}, one can still use supersymmetry parameters to build 
tensors which, in general, are not globally defined
but can be used to define a local 
group structure.
Though this case looses interest from a 
mathematical point of view, it can still be 
very valuable for  
classifying and constructing locally supersymmetric solutions.

\subsection{Static $SU(3)$--structures}
\label{staticsu3}

Let us discuss first the case $G= SU(3)$ for $d=6$ and $d=7$.
The six--dimensional case is well known \cite{chiossi}.
For $d=6$, the generic structure group is $SO(6) \simeq SU(4)$
and the decomposition of $SU(4)$ irrepses under $SU(3)$ gives
\begin{equation}
\label{decomp}
\begin{array}{ccl}
{\bf 4} & \to & {\bf 1}+{\bf 3}\,,\\
{\bf 6} & \to  & {\bf 3} + \overline {\bf 3}\,, \\
{\bf 10} & \to & {\bf 1} + {\bf 3} + {\bf 6}\,,\\
{\bf 15} & \to & {\bf 1} + {\bf 3} + \overline {\bf 3} + {\bf 8} \,.
\end{array}
\end{equation}
This implies the well--known fact that an $SU(3)$ structure in six dimensions is 
specified by an almost complex structure $J$ (and its associated 
2--form) and an invariant complex 3--form $\Psi$, which is of $(3,0)$--type  
with respect to $J$.  These are
the $SU(3)$ singlets of the  corresponding ${\bf 15}$ and ${\bf 
10}$ representations of $SU(4)$. 
In addition, they satisfy the following compatibility relations
\begin{equation}
\Psi \wedge J =0, \;\;\;\;\; \Psi \wedge \overline{\Psi} = -\frac{4 i}
{3} J\wedge J \wedge J.
\label{compat}
\end{equation}
It is also worth noting that (\ref{decomp}) implies the existence 
of two invariant spinors $\eta_{\pm}$.
{From} such spinors (that can be normalized to 1) one can build  
the invariant tensors $J$ and $\Psi$ by contractions with two and 
three gamma matrices respectively.
Then the compatibility relations (\ref{compat}) follow from the 
properties of gamma 
matrices and rearrangements using Fierz identities.

The different $SU(3)$ structures are then classified by the 
decomposition of the torsion $\tau$ into five complex modules
\begin{equation}
\begin{array}{rcccccccccc}
\tau \to \left({\bf 3} + \overline {\bf 3}\right) \times \left({\bf 1} + 
{\bf 3} + \overline {\bf 3}\right) &=& ({\bf 1} + {\bf 1})  \!\!
&+&\!\!\!\!({\bf 8} + {\bf 8})  \!\!
&+& \!\!\!\!({\bf 6} + \overline {\bf 6}) \!\! 
&+& \!\!\!\!({\bf 3} + \overline {\bf 3}) \!\! 
&+& \!\!\!\! ({\bf 3} + \overline {\bf 3})  \!\!\\[2mm]
&=& {\cal W}_1 & +&{\cal W}_2 & +& {\cal W}_3 & + &{\cal W}_4 & + &{\cal 
W}_5\,
\end{array}
\label{eqdeeco}
\end{equation}
and these are completely determined by $dJ$ and $d\Psi$ in the 
following way
\begin{eqnarray}
d J &=& \frac34 \,i\, \left( {\cal W}_1 \,\overline\Psi -\overline{\cal
W}_1\,\Psi\right)  +{\cal W}_3 +  J \wedge {\cal W}_4 \;, 
\label{dJclass}\\[2mm]
d\Psi &=&  {\cal W}_1  J \wedge  J +  J \wedge  
{\cal W}_2 + \Psi\wedge
{\cal W}_5\,, \label{dpsiclass}
\end{eqnarray}
where $ J \wedge {\cal W}_3 =  J \wedge  J \wedge {\cal W}_2 = 0$ 
and $\Psi \wedge {\cal W}_3 = 0$. 
The fact that the $(2,2)$ piece of $d\Psi$ defines the same class as the $(0,3)$
piece of $d J$ is a consequence of the first relation in 
(\ref{compat}).
Operatively one can obtain the various classes by proper contractions 
of $J$ and $\Psi$ with $dJ$ and $d\Psi$. For example,
\begin{equation}
\begin{array}{rclcl}
{\cal W}_1 &=& \displaystyle -{\rm i}\,\frac{3}{32}\,\overline{ \Psi} \,\lrcorner\; dJ  &=& \displaystyle
\frac{1}{12} \,(J\wedge J) \, \lrcorner \; d\Psi\,, \\[4mm]
{\cal W}_4 &=&\displaystyle\frac12 \, J \,\lrcorner\; dJ\,,&
{\cal W}_5 &=\displaystyle\frac14\,\overline{ \Psi} \, \lrcorner\; d\Psi\,.
\end{array}
\label{eqdefalt}
\end{equation}
We remind that a choice of $J$ and $\Psi$ fixes also the 
metric on the six--dimensional manifold and its orientation.
Moreover, one can choose an orthonormal basis of $T^{*}$ such that\footnote{
To avoid cluttering we use the notation ${\rm e}^{i_1\ldots i_n} \equiv
{\rm e}^{i_1}\wedge\cdots\wedge {\rm e}^{i_n}$.} 
$J = {\rm e}^{12}+{\rm e}^{34}+{\rm e}^{56}$ and $\Psi=({\rm e}^1 + {\rm 
i} \,{\rm e}^2)\wedge({\rm e}^3+{\rm i}\,{\rm e}^4)\wedge({\rm e}^5+{\rm i}\,{\rm e}^6)$.


The description of the seven dimensional case is a simple extension 
of the above.
The decomposition of $SO(7)$ to $SU(3)$ gives 
\begin{equation}
\begin{array}{rcl}
{\bf 7}  &\to& {\bf 1}+{\bf 3} + \overline{\bf 3}\,, \\[3mm]
{\bf 21} &\to& {\bf 1}+2 \cdot {\bf 3}+2 \cdot \overline{\bf 3}+{\bf 
8}\,, \\[3mm]
{\bf 35} &\to& 3 \cdot {\bf 1}+2 \cdot {\bf 3}+2 \cdot \overline{\bf 3}+{\bf 
6}+\overline{\bf 6}+{\bf 8}\,.
\end{array}
\end{equation}
Therefore, the only difference with the six--dimensional case is the 
existence of a globally defined vector\footnote{We will use the same symbol
for both the one-form and the dual vector. The precise identification
should be clear from the context.} $v$ (the extra singlet 
3--form is then $J \wedge v$)
An $SU(3)$ structure in $d=7$ is then described by a triplet $v$, 
$J$, $\Psi$, satisfying the compatibility relations (\ref{compat}) and, 
in addition,
\begin{equation}
v\, \lrcorner\; J = 0 \,, \qquad v \,\lrcorner\; \Psi =0\,.
\label{horiz}
\end{equation}
Again, to the $2$--form $J$ one can associate a (1,1)--tensor, which 
now satisfies ${J_a}^b {J_b}^c = - \delta^c_a + v_a \, v^c$.
One can then decompose the horizontal part of the forms according to 
their type with respect to  this tensor.
The vector $v$ allows also for the definition of an almost--product
structure which, if integrable, implies that the metric of the
seven--dimensional space can be written as $ds^{2}_7(x,t) = ds^2_6(x,t)
+ v \otimes v$ with $v = e^{\phi(x)} dt$.

In seven dimensions we have $T^*({\cal M}_7) \otimes SU(3)^\perp
\sim ({\bf 1}+{\bf 3}+\overline{\bf 3}) \otimes ({\bf 1}+2 \cdot {\bf 3}+
2 \cdot\overline{\bf 3})$ and therefore the decomposition of the torsion 
gives a total of 14 classes
\begin{equation}
\begin{array}{rcccccccc}
\tau &\to& 5 \cdot{\bf 1} &+& 4 \cdot({\bf 3} + \overline {\bf 3}) &+& 2\cdot ({\bf 6} + \overline {\bf 
6}) &+& 4\cdot {\bf 8}\,, \\[2mm]
&& R, C_{1,2} &+& V_{1,2}\,,\,W_{1,2} &+& S_{1,2} &+& A_{1,2}\,,\,T\,.
\end{array}
\label{eqclass}
\end{equation}
Notice that $ C_{1,2}$ and $T$ are complex.

Also in this case they can be read from the exterior differentials of 
the forms defining the structure:
\begin{eqnarray}
dv &=& R \,
 J + \overline{W}_1 \lrcorner\; \Psi + 
W_1 \lrcorner\;\overline{\Psi} + A_1  + v \wedge V_1\,,
 \label{dvdef}\\
d J &=& \frac{2 i}{3}\left(C_1\, \Psi - \overline{C}_1\, 
\overline{\Psi}\right) +  J\wedge V_2 
+ S_1 \nonumber \\
&+& v \wedge \left[\frac13 (C_2 + \overline{C}_2)  J + \overline{W}_2 \lrcorner\; \Psi + W_2 \lrcorner\;
\overline{\Psi} + A_2\right] \,,
\label{dJdef}\\
d\Psi &=& C_1  J\wedge J +  J\wedge T + \Psi\wedge
V_3 + v \wedge \left(C_2 \,\Psi -2 J \wedge W_2 + S_2\right). 
\label{dpsidef}
\end{eqnarray}

\subsection{Static $SU(2)$--structures}

The definition of $SU(2)$--structures in six and seven 
dimensions is a bit more involved, but can be obtained from the 
previous one by further decomposing the $SU(3)$ 
representations in terms of $SU(2)$ ones.

Using that ${\bf 3} \to {\bf 1} + {\bf 2}$, ${\bf 6} \to {\bf 1} + {\bf 
2} + {\bf 3}$ and ${\bf 8} \to {\bf 1} + {\bf 2} + {\bf 2} + {\bf 
3}$, it follows that an $SU(2)$ structure in six dimensions is specified 
by an invariant complex 1--form $w$, one invariant 2--form $J$, and 
one invariant complex 2--form $K$. 
All the extra singlets in the $SU(2)$ decomposition can be written in 
terms of these three objects.
Compatibility of these forms now imposes that \cite{Gauntlett:2003cy}
\begin{equation}
\label{su2comp}
K\wedge K =0, \quad J \wedge K =0, \quad K \wedge \overline{K} = 2 \,J\wedge 
J \,,
\end{equation}
as well as
\begin{equation}
w \lrcorner\; K =\overline{w} \lrcorner\; K = 0, \quad
w \lrcorner\; J =\overline{w} \lrcorner\; J = 0.
\end{equation}
In what follows we will also often further decompose real and 
imaginary parts as
\begin{equation}
K \equiv J_2+ {\rm i} \,J_3\,, \quad w \equiv w_1 + {\rm i} \,w_2\,,
\label{eqKw}
\end{equation}
and define $J_1 \equiv J$.

Locally one can introduce a frame such that:
\begin{equation}
\label{eq:su2frame}
w={\rm e}^5+i \,{\rm e}^6,\;\;\;
J={\rm e}^{12}+{\rm e}^{34}, \;\;\; K=({\rm e}^1+ {\rm i} \,{\rm e}^2)\wedge 
({\rm e}^3+ {\rm i}\, {\rm e}^4).
\end{equation}
Notice that $J$ can be thought as an almost complex structure
in the 4--dimensional part of the tangent bundle spanned by 
$\{{\rm e}^i\},i=1,\ldots,4$, with respect to which $J$ is of $(1,1)$ and
$K$ of $(2,0)$ type.
In the same way, the triplet of two--forms $J_i$ induce a triplet of 
almost--complex structures satisfying $J_i J_j = - \delta_{ij} + 
\epsilon_{ijk} J_{k}$. 
 
Now $T^*({\cal M}_6) \otimes SU(2)^\perp
= \left(2\cdot {\bf 1} +2 \cdot {\bf 2} \right) \otimes \left(
4\cdot {\bf 1}+4\cdot {\bf 2}\right)$ and therefore the decomposition of the torsion 
gives a total of 20 classes
\begin{equation}
\begin{array}{rcccccc}
\tau & \to & 16\cdot {\bf 1} &\oplus& 16\cdot {\bf 2}&\oplus& 8 \cdot {\bf 3}\,  \\
 &  & S_{1,\ldots,8} && V_{1,\ldots,8}&& T_{1,\ldots,4}\,.
\end{array}
\end{equation}
As usual, one can define these 20 classes from the exterior 
differentials on the forms defining the $SU(2)$ structure
\begin{equation}
\begin{array}{rcl}
dw & = & S_1\, K + w \wedge V_1 + 
S_2\, J + \overline{w} \wedge V_2 +
S_3\, w \wedge \overline{w}  + T_1 
+ S_4\, \overline{K}\,, \\[2mm]
dJ &=&\displaystyle S_5\, \left(K \wedge w \right)+
S_6\, \left(K \wedge \overline{w}\right)  + \frac12(S_7 + \overline{S}_8)\, J 
\wedge w  + J \wedge V_4\,, \\[2mm]
& +& w \wedge 
\overline{w} \wedge V_5 + w \wedge T_2 +
c.c \\[2mm]
dK&=& S_7\, K \wedge w + S_8\, K\wedge\overline{w}  -2 \overline{S}_5\, 
J \wedge 
w  + J \wedge V_6 + {\rm i}\,
w \wedge\overline{w} \wedge (\overline{V}_5 \lrcorner K) \\[2mm]
& +& w \wedge T_3 - 2 \overline{S}_{4}\,  J  \wedge \overline{w}  + J \wedge 
\overline{V}_8 
+\overline{w} \wedge T_4\,.
\end{array}
\end{equation}
Here the torsion components satisfy the following consistency relations
\begin{equation}
J \wedge T_i = K \wedge T_i = 0, \;\;\; K \wedge V_i = 0, \;\;\; J\wedge J\wedge V_i=0.
\end{equation}


The $SU(2)$ structures in seven dimensions are now straightforward to 
obtain. 
One has simply an extra globally defined vector $v$.
To keep the notation compact we denote 
$\{v,w_1,w_2\}$ collectively by $v^i$, with $i =1,2,3$.
The intrinsic torsions of the $SU(2)$ structure are then
\begin{equation}
\begin{array}{rcl}
dv^i &=& C^{ij} J^j + S^{ij} \,\epsilon_{jkl} \,v^k \wedge v^l + 
{W^i}_{j}\wedge v^j+ T^i\,, \\[2mm]
dJ^i &=& C^{ijk} J^j \wedge v^k + {\epsilon^i}_{jk}\, J^j \wedge W^k + 
V^{ij} \wedge v^k \wedge v^l \,\epsilon_{jkl} + {T^i}_{j} \,v^j\,,
\end{array}
\end{equation}
where
\begin{eqnarray}
C^{ijk}=\delta^{ij}\hat{C}^k+\epsilon^{ijm}\,\hat{C}_{m}^k\,,
\;\;\;
J^1\wedge V^{1k} =  J^2\wedge V^{2k} = J^3\wedge V^{3k}\,,
\end{eqnarray}
due to the consistency conditions (\ref{su2comp}), to which one has to 
add $v \, \lrcorner\; J^i = 0$.
The number of independent classes is 30 singlets, 15 doublets along
with their conjugates and 30 triplets of $SU(2)$, exactly as expected
from $T^*({\cal M}_7) \otimes SU(2)^\perp$.

\section{Intrinsic torsion classes for M--theory with fluxes}
\label{Mthconds}

In this section we derive the necessary and sufficient 
conditions for obtaining supersymmetric solutions of M--theory with fluxes. 
In doing so, we will show how to make contact with \cite{Kaste:2003zd}, 
where necessary conditions were found and discussed in terms of 
$G_2$--structures.

The gravitino variation of eleven--dimensional supergravity in the 
presence of a nontrivial 4--form flux $G=dC$ and with vanishing
gravitino background values reads 
\begin{equation}
\delta \Psi_A = \left\{ D_A[\omega] +\frac{1}{144} G_{BCDE}
 \left(\Gamma^{BCDE}{}_A - 8 \Gamma^{CDE} {\eta^B}_A \right) 
 \right\} \epsilon , \label{grav.flux.var.11}
\end{equation}
where $\epsilon$ is a Majorana spinor in eleven dimensions and we have
denoted flat indices by using letters from the beginning of the 
alphabet and curved ones using letters from the middle of the alphabet.

In what follows we consider warped compactifications to 
four--dimensional Minkowski space--time   
\begin{equation}
ds_{11}^2 = e^{2\Delta}\, \eta_{\mu \nu} dx^{\mu} dx^{\nu} + ds_7^2 \,, \label{ds11.w}
\end{equation}
where the warp factor depends only on the internal coordinates, 
$\Delta=\Delta(y^m)$.
Now greek letters will be used for the 4d part and small latin 
letters for the internal manifold.
Poincar\'e invariance of the four--dimensional 
part of the solution allows a non-zero four--form flux $G$ 
only on the internal manifold and  depending only on the internal coordinates.
We are not going to discuss modifications due to a non--trivial 
cosmological constant in space--time, since this is a simple 
extension \cite{Kaste:2003zd,Behrndt:2003uq}.

The decomposition for the eleven--dimensional $\gamma$--matrices is the standard one
\begin{equation}
\Gamma^{\alpha} = \gamma^{\alpha} \otimes \mathbb I \,, \quad
\Gamma^a = \gamma^{(5)} \otimes \gamma^a \,,
\end{equation}
where $\gamma^{(5)}=i\gamma^1\gamma^2 \gamma^3\gamma^4$ is the
four--dimensional chirality operator.  
A useful choice for these matrices is given by the Majorana
representation.
In this representation the $\gamma$--matrices are either purely real
($\gamma^{\alpha}$) or purely imaginary ($\gamma^{(5)}$ and
$\gamma^a$) and the Majorana condition on $\epsilon$ reduces to
the reality constraint $\epsilon^*=\epsilon$.
One can therefore split the supersymmetry parameter as
\begin{equation}
\epsilon=\psi_+ \otimes \eta_+ + \psi_- \otimes \eta_-  \label{ans.eps}
\end{equation}
where $\eta_{\pm}$ depend only on the internal coordinates
and the $\pm$ label refers to the chirality of the four--dimensional 
part. 
The Majorana 
constraint on $\epsilon$ then requires $(\psi_{\pm})^*=\psi_{\mp}$ and 
$(\eta_{\pm})^*=\eta_{\mp}$.

The gravitino variation 
(\ref{grav.flux.var.11}) leads to the following supersymmetry
constraints on the internal spinors:
\begin{equation}
\left[ \pm  \frac{1}{2} (\partial_c \Delta) \gamma^c 
 + \frac{1}{144} 
  G_{bcde} \gamma ^{bcde} \right] \eta_{\pm}=0\,, \label{var.const.1}
\end{equation}
from the space--time part $\alpha=0,\ldots,3$ and
\begin{equation}
\begin{array}{rcl}
D_a[\omega] \eta_{\pm} &=& \displaystyle
  \mp \frac{1}{144}
 \left( G_{bcde} \gamma ^{bcde}{}_a - 8 G_{abcd} \gamma ^{bcd}
   \right)  \eta_{\pm} \nonumber\\[4mm]
&=&  \displaystyle 
 \pm  \left( \frac{i}{12} (*G)_{abc} \gamma^{bc} 
 + \frac{1}{18} G_{abcd} \gamma ^{bcd}
   \right)  \eta_{\pm}  \label{var.const.3}
\end{array}
\end{equation}
from the internal part $a=4,\ldots,10$. 
We have 
defined $(*G)_{abc}\equiv \frac{1}{4!}\epsilon_{abcdefg}G^{defg}$.


The existence of $\eta_{\pm}$ implies the definition of an $SU(3)$ 
structure since they are in one to one correspondence with the two 
singlets of the decomposition of the fundamental representation of $Spin(7)$.
In order to discuss the $SU(3)$ structure as in  section 
\ref{staticsu3}, one first needs to normalize them properly.
This normalization can be read from (\ref{var.const.3}) by considering 
the contraction of that equation with $\eta^{\dagger}_{\pm}$.
Defining\footnote{It can be shown that $\eta^{\dagger}_- \eta_+$ vanishes because $\eta_{\pm}$ are 
orthogonal. If this does not happen then one has just one independent 
supersymmetry parameter and non--trivial 4--form flux necessarily curves 
the space--time \cite{Behrndt:2003uq}.} 
$\Xi \equiv \eta^{\dagger}_+ \eta_+ = \eta^{\dagger}_- \eta_-$, one obtains 
\begin{equation}
d\left(e^{-\Delta} \Xi\right) = 0\,,
\label{eqdXi}
\end{equation}
which implies that a good normalization is given by $\Xi = e^{\Delta}$.

The three tensors  defining the $SU(3)$ structure are then obtained 
from
\begin{equation}
\begin{array}{rcl}
v_{a} & = & e^{-\Delta}\, \eta^{\dagger}_+ \gamma_a \eta_+\,,  \\[2mm]
J_{ab} & = & - {\rm i} \,e^{-\Delta}  \,\eta^{\dagger}_+ \gamma_{ab}\eta_+\,,  \\[2mm]
\Psi_{abc} & =  & - {\rm i}\, e^{-\Delta}\, \eta^{\dagger}_- \gamma_{abc} \eta_+\,.   
\end{array}
\label{eqdefsu3}
\end{equation}
By using the gamma--matrices relations, one can indeed check 
that such tensors (and the corresponding forms) 
satisfy the requirements of section \ref{staticsu3}.
Moreover, it can be shown that these are the only independent 
contractions of the gamma--matrices with the $\eta_{\pm}$ spinors since
\begin{equation}
\begin{array}{rclcl}
\eta^{\dagger}_- \gamma_{[n]} \eta_+ &=& 0\,, &\quad& \hbox{for } 
n=0,1,2\,,\\[2mm]
e^{-\Delta}\, \eta^{\dagger}_+ \gamma_{abc}\eta_+ &=& 3\,{\rm i}  \,
v_{[a}J_{bc]}\,,
\end{array}
\label{eqrelind}
\end{equation}
and the rest follow from duality relations.

\medskip

We are now ready to interpret the supersymmetry conditions 
(\ref{var.const.1}) and (\ref{var.const.3}) as conditions on the 
$SU(3)$ structure of the internal manifold defined by (\ref{eqdefsu3}).
In order to do so, one considers the contraction of these two equations 
with the full basis of tensor--spinors constructed from 
$\eta^{\dagger}_{\pm} \gamma^{[n]}$.
Then, using some gamma--algebra and the definitions of the $SU(3)$ 
structure tensors, the independent conditions will be summarized as 
constraints on the torsion classes and on the allowed fluxes.
The last ingredient one needs is the decomposition of the four--form 
flux in terms of irreducible $SU(3)$ representations.
Since it is a four--form it decomposes as ${\bf 35} \to 3 \cdot{\bf 1} \oplus 
2\cdot ({\bf 3}\oplus \overline{\bf 3}) \oplus ({\bf 6} \oplus \overline{\bf 6}) 
\oplus {\bf 8}$ and one can therefore write
\begin{equation}
G= -\frac{Q}{6}\, J \wedge J  +  J \wedge A + \psi_- \wedge V+ v \wedge 
\left(c_1 \, \psi_+ + c_2 \, \psi_- + J \wedge W + U\right)\,.
\label{eqGdeco}
\end{equation}
Here the first term was normalized in a way that the singlet $Q$ 
corresponds to the same one shown in \cite{Kaste:2003zd} in the $G_2$ 
decomposition of $G$ for supersymmetric configurations.
The same thing can be done for the dual, which then reads\footnote{
To be consistent with \cite{Kaste:2003zd} and in order to compare  
results, we had to choose the volume form as $ -\frac16 J\wedge J 
\wedge J \wedge v$, so that for 
instance $* J = -\frac12 \,J \wedge J \wedge v$.}
\begin{equation}
*G = \frac{Q}{3} \, J \wedge v + v \wedge A - J \wedge\left( W\lrcorner\; 
J \right) + 
S + c_1' \,\psi_+ + c_2' \, \psi_- + v \wedge \left(V \lrcorner\; \psi_+\right)\,.
\label{eqstarG}
\end{equation}
We emphasize that $*$ denotes the 7-dimensional Hodge duality operator.

Since $A$, $S$, $V$, $U$ and $W$ are horizontal with respect to $v$, 
one can decompose them according to their type with respect to $J$.
For instance $A$ is a primitive $(1,1)$--form, $S$ is a primitive 
$(2,1)$--form plus its complex conjugate and so on.
For future reference, we also give the definition of some 
components of the flux as contractions with the structure tensors
\begin{equation}
\begin{array}{rclcrcl}
Q & = &\displaystyle-\frac12\, \left(J\wedge J\right)\,\lrcorner\; G\,,  
&\quad&
W  & = & \displaystyle \frac12 \,\left(J\wedge v\right) \, \lrcorner\; 
G\,,   \\[2mm]
A & = & \displaystyle J \, \lrcorner\; G + \frac23 Q \, J - 2 \, v \wedge W\,, 
&\quad&
U &=& v \,\lrcorner \; G - J \wedge W\,.   
\end{array}
\label{eqcompo}
\end{equation}

The first condition following from supersymmetry, namely 
(\ref{var.const.1}), does 
not contain derivatives of the spinors and therefore will be realized 
as simple constraints on the flux and an equation for the warp factor.
The independent conditions on the flux are
\begin{equation}
\Psi \lrcorner\; G = 0 = \overline{\Psi} \lrcorner\; G\,,
\label{eqGcond}
\end{equation}
which remove two singlets and one vector from (\ref{eqGdeco}) and (\ref{eqstarG}).
The other independent conditions can be written as an equation for 
the warp factor
\begin{equation}
d\Delta = -\frac13 Q \, v + \sigma\,,
\label{eqdDelta}
\end{equation}
and a further relation between $\sigma$ and the remaining one--form 
in the fluxes:
\begin{equation}
\sigma = \frac23\, W \lrcorner\; J\,.
\label{eqsgm}
\end{equation}
It should be noted that there are no constraints at all concerning the 
primitive (1,1)--form $A$ and the (2,1)--form $S$. 

After some tedious calculations one can also obtain differential 
conditions on the $SU(3)$ structures by using (\ref{var.const.3}).
The resulting conditions can be summarized in the following concise 
expressions\footnote{These relations and the above conditions on the flux are in agreement 
with \cite{Kaste:2003zd}, but for a typo in their eq. (3.19) where the 
r.h.s. should have a factor of 2. The conditions on the flux also give 
naturally $Q_{ab} v^b = 0$ and not the weaker condition 
$Q_{ab}v^a v^{b}= 0$, but, as noted 
also by A. Tomasiello, using 
the expression for $Q_{ab}$ in terms of its irreducible components and 
the other conditions on the flux one can show that they are equivalent.}:
\begin{eqnarray}
dv & = & 2 \, v\wedge d\Delta\,, \label{eqdv}\\[2mm]
dJ & = & -4\, J \wedge d\Delta - 2 * G\,, \label{eqdJ}\\[2mm]
d\Psi & = & 3\, \Psi \wedge d\Delta\,. \label{eqdpsi}
\end{eqnarray}
From (\ref{eqdJ}) we conclude that $J$ is a generalized calibration
(c.f. section 4.1.2) of the 7-dimensional internal space.

Using the expression for the dual flux (\ref{eqstarG}), the equation 
for the warp factor (\ref{eqdDelta}) and the relation (\ref{eqsgm}) 
one can further rewrite (\ref{eqdv})--(\ref{eqdpsi}) completely in terms 
of the $G$--flux components.
In detail, it is easy to check that 
\begin{eqnarray}
dv & = & 2 \, v\wedge \sigma \,, \label{eqdv2}\\[2mm]
dJ & = & \frac23\, Q \,v \wedge J - 2 S - 2 \, v \wedge A - J \wedge \sigma   
\,, \label{eqdJ2}\\[2mm]
d\Psi & = & -\,Q\, \Psi \wedge v + 3\, \Psi \wedge\sigma \,. \label{eqdpsi2}
\end{eqnarray}
A comparison with (\ref{dvdef})--(\ref{dpsidef}) now yields 
\begin{equation}
R = C_1 = W_1 = W_2=A_1 = T = S_1 = 0\,,
\label{eqzero}
\end{equation}
and the following identifications 
\begin{equation}
\begin{array}{rclrcl}
C_2 & = & \overline C_2 = Q\,, & V_1 & = & \displaystyle \frac23 V_3 = \sigma\,,  
\\[3mm]
A_2 & = & -A\,, & S_1 & = & -2S\,.   
\end{array}
\label{eqrels}
\end{equation}
It should also be noted that consistency requires some differential 
constraints on these torsion components:
\begin{eqnarray}
&&3 \, d\sigma = dQ \wedge v + 2 Q\, v \wedge \sigma\,,
\label{constr0}\\[2mm]
&&v \wedge \Psi \wedge \left( dQ - 2 \,Q\, 
\sigma\right) = 0\,,
\label{constr1}\\[2mm]
&&3 \, d\sigma \wedge J - 6 \,dS + 6\, v \wedge dA=  - 4 Q \, v \wedge 
S + 6\, v \wedge \sigma \wedge A - 6 \,S 
\wedge \sigma\,.
\label{constr2}
\end{eqnarray}

Supersymmetric backgrounds for M--theory with fluxes can now be 
constructed by choosing seven--dimensional manifolds with an $SU(3)$ 
structure whose torsion classes are of the form (\ref{eqrels}).
Of course, one has to check the equations of motion for the 
four--form field $G$ and the metric, and also the Bianchi identity for 
$G$.
However, as proven in \cite{Gauntlett:2002fz}, if one solves the 
supersymmetry conditions, the Bianchi identity for $G$ and its 
equation of motion, the Einstein equation is identically satisfied.
Let us then analyze these conditions.

The equation of motion for the four--form can be written in terms of 
differential forms as
\begin{equation}
d \left(*_{11} G\right) = G \wedge G\,,
\label{eqeomG}
\end{equation}
where the Hodge dual is taken with respect to the full 
eleven--dimensional metric.
Since we are considering configurations admitting only a 
non--vanishing expectation value for $G$ in the internal space, one 
can rewrite (\ref{eqeomG}) as
\begin{equation}
d \left(*\, e^{4 \Delta}\,  G\right) = 0\,,
\label{eomG7}
\end{equation}
where now the Hodge dual is taken only with respect to the internal 
space and the term $G \wedge G$ is vanishing because it is an 
8--form in 7--dimensions.
It is then straightforward to check that such an equation is 
identically satisfied for our backgrounds \cite{Becker:2000rz}, 
as (\ref{eqdJ2}) implies that $*\, e^{4 \Delta}\,  
G =-\frac12  d\left(e^{4\Delta}\,J\right)$.
This finally means that the complete equations of motion are satisfied 
once the supersymmetry conditions (\ref{eqdv2})--(\ref{eqdpsi2}) are 
fulfilled along with the Bianchi identity $dG =0$. We should mention
that we consider for the moment the source-free Bianchi identity. Later on
we will have to relax this restriction and allow for the possibility
of (wrapped) M5--branes which will result in a non-zero contribution to
$dG$.

In conclusion, in order to get a complete 
set of necessary and sufficient conditions for supersymmetric 
solutions of M--theory with fluxes, one just needs to understand the 
4--form Bianchi identity.
Assuming that there are no sources, one can obtain 
differential conditions on the flux components
from the $SU(3)$ decomposition of the $G$--flux
by requiring that $dG = 0$.
These conditions read
\begin{eqnarray}
\frac{dQ}{6}\wedge J \wedge J + \frac{2}{9} Q^2 \, J\wedge J\wedge v  - \frac{4}{3} Q 
\,v\wedge J\wedge A - 
\frac{Q}{3} J\wedge J \wedge \sigma +
 2 \,S\wedge A + 2\, v\wedge 
A\wedge A  &&\nonumber \\[2mm]
+ J \wedge \sigma\wedge  A - 3\, v\wedge \sigma\wedge 
J\wedge 
W - 2\, v\wedge S\wedge W + v\wedge J\wedge dW + d*S = 0\,.
\label{eqdG}
\end{eqnarray}
As it is obvious, this is not a simple restriction on the torsion 
classes, but it imposes rather non--trivial differential relations. 
We will see in the construction of explicit examples that such 
relation can result in reasonable restrictions on the fluxes and the 
geometry and that it can be identically satisfied in certain cases.
Of course, a non-vanishing $dG$ will not be disastrous, provided that
it can be interpreted as due to the presence of 5--brane sources.

\section{Irreducible 7--manifolds}

The first type of solutions we want to discuss are 7--manifolds ${\cal 
M}_7$ which do not admit any isometries and therefore we will call  
them irreducible.
In detail, we are going to focus on fibered products of 6--manifolds 
${\cal M}_6$
with an interval $I \subset  \mathbb{R}$. 
Moreover, we assume ${\cal M}_6$ allows for an $SU(3)$ structure 
$\{\widehat{J},\widehat{\Psi}\}$ which will also be fibered over $I$.
This means that there is a naturally induced $SU(3)$--structure on the 
7--dimensional manifold given by $\{v =e^{q 
\phi}dt,J=\widehat{J}(t),\Psi=\widehat{\Psi}(t)\}$, where $t$ is the 
variable parameterizing $I$.

For the sake of clarity we discuss the case $q=0$ separately than
the general case; as we will see this distinction is actually 
physically relevant, in the sense that the two classes of 
solutions we obtain
admit different physical interpretations. Moreover, the case $q=0$
provides a direct generalization of the Hitchin construction of 
$G_2$--holonomy manifolds.

\subsection{Case I: $q=0$}
\label{irredI}

In this case, the metric of the 7--dimensional
spaces reads
\begin{equation}
ds^2_{7}(y,t)= ds^2_{6}(y,t) + dt^2\,.
\end{equation}
As previously assumed, at any given $t$, the 6--dimensional manifold 
has an $SU(3)$ structure.
This cannot be generic if we require a supersymmetric 
solution of M--theory.
In detail, its structure should follow from the restriction of 
(\ref{eqdv2})--(\ref{eqdpsi2}) for a given $t$.
In order to understand this, it is useful to split the 7--dimensional 
differential $d$ as $d=\widehat{d}+ dt\,\frac{\partial}{\partial t}$, 
$\widehat d$ being the 6--dimensional one.
Using this fact and the above definition of the 7--dimensional 
$SU(3)$ structure we can finally provide the 6--dimensional structure 
and a set of differential equations in $t$ which describe the 
fibration of this structure over $t$.
Before giving these conditions explicitly let us note that 
since $dv =0$ by construction, the vector component of the flux has to 
vanish: $\sigma=0$. 

The conditions on ${\cal M}_6$ read
\begin{equation}
\left\{
\begin{array}{rcl}
\widehat{d}J&=&-2 S\,, \\
\widehat{d}\Psi&=&0\,,
\end{array}
\right.
\label{specialhermitian}
\end{equation}
and this means that ${\cal M}_6$ is a special--hermitian 
manifold, i.e. it is a complex, non--K\"aher manifold with 
$\widehat{d} J \neq 0$ but $\widehat{d} \left(J\wedge J\right) = 0$.
Given such type of manifolds, one can build a 7--manifold that can be 
used in a flux solution of M--theory, by solving the following first 
order differential equations in $t$:
\begin{equation}
\label{flow}
\left\{
\begin{array}{rcl}
\displaystyle\frac{\partial J}{\partial t} &=&\displaystyle 
\frac{2}{3} \, Q \, J - 2 A\,, \\[3mm]
\displaystyle\frac{\partial \Psi}{\partial t} &=& Q \,\Psi \,.
\end{array}
\right.
\end{equation}
This construction mimics the Hitchin construction of $G_2$--holonomy 
manifolds \cite{Hitchin}
as presented in \cite{chiossi}.
In that case the base is a half--flat manifold\footnote{Recall that
a 6-dimensional manifold with an $SU(3)$ structure is called half-flat
when $d\psi_+ =0$ and $J\wedge dJ=0$.} 
and the flow equations are
$\partial_t \psi_+ = \widehat d J$, $\partial_t (J \wedge J) = -2 \widehat d 
\psi_-$, where the split $\Psi = \psi_+ + {\rm i}\, \psi_-$ is used.
It should also be noted that (\ref{flow}) preserve the compatibility relations 
$\widehat{\Psi} \wedge \widehat{J} =0$ and $\widehat{\Psi} \wedge 
\overline{\widehat{\Psi}} = -\frac{4 i}
{3} \widehat{J}\wedge\widehat{J}\wedge\widehat{J}$ .
It is tempting to conjecture that for the common class of 
6--dimensional base spaces given by special--hermitian manifolds the two 
pictures are related. 
The idea would be that by turning on an appropriate flux on 3--cycles 
of ${\cal M}_6$ one could obtain a deformation of its fibration, 
described by the set of equations just presented. 
We present a very simple example along these lines in the next
subsection, but it would be interesting to make this idea more precise.

Given this general construction, let us now present some explicit 
examples in order to illustrate better the procedure one has to follow.
We stress that even though we specify the flux by solving the conditions 
rather than trying to obtain solutions for a given flux, the procedure 
is general.

\subsubsection{Calabi--Yau base}

As a first instance of special hermitian 6--manifolds we analyze the 
case $S=0$, i.e. ${\cal M}_6$ is Calabi--Yau $(\widehat d \widehat J = 0 = \widehat 
d \widehat \Psi)$.
In order to further simplify the setup, we also demand that the flux 
depends only on $t$ and therefore $\widehat{d}Q=0$.
{From} the first flow equation follows $\widehat{d} A=0$ and the
Bianchi identity  (\ref{eqdG}) for $G$ becomes
\begin{equation}
-\frac{1}{6}\left(\dot{Q}+\frac{4}{3}Q^2\right) \,J\wedge J + \frac{2}{3}
Q \,J\wedge A-2 A\wedge A\ + J\wedge \dot A = 0\,,
\end{equation}
which can be easily solved by setting $A=0$.
In this case one has to satisfy $\dot{Q}+\frac43 Q^2 =0$, which 
specifies
\begin{equation}
Q(t)=\frac{3 q_{0}}{3+4 q_0 t} \,,
\end{equation}
where $q_0$ will be associated with the ``charge" of the solution.
With this information, the warp factor is also completely determined 
by (\ref{eqdDelta})
\begin{equation}
e^{\Delta(t)}=\left(1+\frac{4 q_0}{3} t\right)^{-\frac{1}{4}} \,,
\end{equation}
where we chose $\Delta(0)=0$ for simplicity.

The $SU(3)$ structure in 7 dimensions is now easily fixed by using 
(\ref{flow}). All in all, the $t$--dependence results in an overall factor in 
front of the 6--dimensional structure:
\begin{eqnarray}
J(t)&=&\left(1+\frac{4 q_0}{3} t\right)^{\frac{1}{2}} \widehat{J}\,, \\
\Psi(t)&=&\left(1+\frac{4 q_0}{3} t\right)^{\frac{3}{4}} 
\widehat{\Psi}\,.
\end{eqnarray}
The full metric of 11--dimensional supergravity is then given by
\begin{equation}
ds^2_{11}(x,y,t)= \frac{1}{\sqrt{1+\frac{4 q_0}{3} t}}\;
\eta_{\mu \nu} dx^{\mu} dx^{\nu}
+\sqrt{1+\frac{4 q_0}{3} t}\;
ds^2_{CY}(y) + dt^2
\label{heteroticM}
\end{equation}
where $ds^2_{CY}(y)$ is a 6-dimensional Calabi--Yau metric. 

The flux is completely determined by its singlet component 
$Q$ and it reads
\begin{equation}
G=-\frac{Q}{6} J\wedge J = -\frac{q_0}{2} \widehat{J}\wedge\widehat{J}.
\end{equation}
One can therefore interpret this solution as a deformation of the 
direct product of Calabi--Yau manifolds with $I$, which corresponds
to $q_0 =0$, when a 
non--trivial flux of strength $q_0$ is turned on
the 4--cycles of the Calabi--Yau.

Such a solution can be employed as a genuine heterotic--M--theory 
background once the space--time fields satisfy appropriate conditions 
at the boundaries of $I$ \cite{Curio:2003ur}.
The metric (\ref{heteroticM}) corresponds to an isotropic deformation 
of the CY metric over the interval $I$.
It would be very interesting to construct anisotropic solutions by allowing 
for a non--zero $A$.

\subsubsection{M5--branes in special--hermitian manifolds}

We now discuss generic flux configurations based on special--hermitian
manifolds with M5--branes; as a concrete application
we analyze the case of M5--branes wrapping 2--cycles
of the Iwasawa manifold.

Our starting point is arbitrary fibrations of special--hermitian manifolds ${\cal M}_6$
over the interval $I$, with fluxes satisfying (\ref{specialhermitian}) and
(\ref{flow})
\begin{eqnarray}
S&=&-\frac{1}{2} \widehat{d}J \\
A&=&\frac{1}{3} Q J -\frac{1}{2} \dot{J} .
\end{eqnarray}
Notice that this choice implies identically 
the constraint $\widehat{J}\wedge\widehat{J}\wedge A=0$.
In addition, 
the exterior differential of $G$ reads
\begin{equation}
dG=-\frac{1}{6}\left(\dot{Q}+\frac{4}{3}Q^2\right) {J}\wedge{J}\wedge dt + \frac{4}{3}
Q \,{J}\wedge A\wedge dt -2 A\wedge A\wedge dt + {J}\wedge \dot{A}\wedge
dt + \widehat{d} (* S),
\end{equation}
which will not vanish in general. Hence, most of these solutions will be
interpreted as M5--branes wrapping 2--cycles in ${\cal M}_7$.

Before we discuss wrapped M5-branes it is useful to review first 
some facts about generalized calibrations.
Recall that a
generalized calibration of degree $k$ in a manifold ${\cal M}$ 
is a $k$-form $\phi$ such that its pull-back on any $k$-dimensional submanifold
$\Sigma$ of ${\cal M}$ is less than or equal to the corresponding volume form, i.e.
\begin{equation}
\phi^* \leq {\rm vol}_{\Sigma} .
\end{equation}
As opposed however to standard
calibrations,
$\phi$ does not have to be a closed form. A submanifold ${\cal N}$ of ${\cal
 M}$ is called calibrated with respect to $\phi$ if $\phi^* = {\rm vol}_{\cal
 N}$. 

The relevance of generalized calibrations to branes in
curved backgrounds with fluxes is due to the fact that submanifolds
calibrated by $\phi$ minimize functionals of the form $E(\Sigma)=\int_{\Sigma}\left(
{\rm vol}_{\Sigma}- A\right) $, with $A$ a $k$-form such that $d\phi=dA$.
These functionals correspond precisely to the energy\footnote{When $k<p$ we
  should actually talk about the energy density of the wrapped brane
since the energy of an infinitely extended brane is infinite.}
of a $p$-brane wrapping
a $k$-dimensional submanifold of spacetime in the presence of background fluxes;
the first part is the standard worldvolume contribution, 
while the second originates
from the Wess-Zumino term in the brane action which
couples the brane with the appropriate spacetime form--field.
The calibrated submanifold minimizes $E(\Sigma)$ among the homology
class of $\Sigma$. Notice however that unlike the case of standard
calibrations, the cycles corresponding to generalized calibrations may also be
topologically trivial. For more details on generalized calibrations
consult \cite{Gutowski:1999iu,Gutowski:1999tu,Barwald:1999ux,Gutowski:2002bc}.

In general, branes wrapping calibrated submanifolds are
BPS and they preserve some supersymmetry. The relevant conditions
on the fluxes are precisely of the form $d\phi=dA$, where $\phi$ is a purely
geometrical object, and they determine the corresponding generalized
calibration.
For example, in the situation described above we are interested in 
M5--branes wrapping a 2-dimensional submanifold of ${\cal M}_7$
in the presence of 4-form flux $G$. Since the M5--branes couple magnetically
to $G$, the generalized calibration satisfies
\begin{equation}
d \phi=*_{11}G,
\end{equation}
which, compared to the condition (\ref{eqdJ}) coming from 
supersymmetry, implies that
\begin{equation}
\phi =  -\frac{1}{2}e^{4 \Delta} dx^0 \wedge dx^1 \wedge dx^2 \wedge dx^3
\wedge J .
\end{equation}
This is expected since the full calibrating form should consist 
of the volume
form of the flat part of the brane\footnote{Notice that due to the warp factor that volume form
of the flat 4-dimensional part is actually
$ e^{4 \Delta} dx^0 \wedge dx^1 \wedge dx^2 \wedge dx^3$.} 
along with a (generalized) calibration of degree 2 given by $J$. 
The fact that $J$ is indeed a generalized calibration 
is just a fiber-wise application of the fact that 
$\widehat{J}$ is a generalized calibration on any almost-hermitian (and
consequently on any special--hermitian) manifold \cite{Gutowski:2002bc}.
Notice that the above calibrating form is the most general possible, 
but in actual examples one can consider its restriction to a specific 
configuration of calibrated cycles.

It is
important to mention 
that a priori the flux $G$ may have nothing to do with the 5--branes,
i.e. can be thought of as a background 4-form flux. 
If we want however to interpret our solution as a configuration of 
wrapped M5--branes with no other background fluxes, we have to identify
$G$ with the flux due to the branes. Then we obtain a further relation between
the calibration and the flux, this time coming from the
Bianchi identity with magnetic sources
\begin{equation}
\label{eq:bianchi}
dG=*_{11} {\cal J}_6 ,
\end{equation}
with ${\cal J}_6$ a 6-form specifying where the 5--branes are located in
the transverse 5--dimensional space and how they are oriented.


We present now a concrete example of this type based on
the Iwasawa manifold \cite{AGS,Cardoso:2002hd}.
This manifold is a $\mathbb{T}^2$ fibration over  $\mathbb{T}^2 \times
\mathbb{T}^2$ and its  6--dimensional orthonormal frame ${\rm e}^i,i=1,\ldots,6$  satisfies
\begin{equation}
\left\{
\begin{array}{rcl}
d {\rm e}^i &=& 0, \; i = 1,\dots,4,  \\[2mm]
d {\rm e}^5 &=& {\rm e}^{13}- {\rm e}^{24}, 
\;\;\; \\[2mm]
d {\rm e}^6 &=& {\rm e}^{14}+ {\rm e}^{23}. 
\end{array}     
\right.
\end{equation}  
In terms of complex coordinates we have
\begin{equation}
dz={\rm e}^1+{\rm i}\, {\rm e}^2,\;\;\;dv={\rm e}^3+{\rm i}\, {\rm e}^4,\;\;\;
-du+z dv={\rm e}^5+{\rm i}\,{\rm e}^6 \,.
\end{equation}

Given $\widehat J$ the canonical choice of complex structure on the 
Iwasawa manifold, i.e.
\begin{equation}
\widehat{J}\equiv {\rm e}^{12}+{\rm e}^{34}+{\rm e}^{56}=\frac{i}{2}\left(dz \wedge d{\overline z}+dv
  \wedge d{\overline v}+(-du+z dv) \wedge (-d{\overline u}+{\overline z} d{\overline v})\right) ,
\end{equation}
it can be checked that
\begin{equation}
\widehat{d}\widehat{J}=
\left({\rm e}^{136}-{\rm e}^{246}-{\rm e}^{145}-{\rm e}^{235}\right)\,.
\end{equation}
We are therefore interested in fibrations of such a manifold with a 
real interval $I$.
The uplift of the complex structure $\widehat J$ to the full 7--dimensional 
space determines also the metric of such space and can be rather 
general.

As a working assumption, we discuss the case of simple size 
deformations:
\begin{equation}
J = e^{2 a(t)} {\rm e}^{12}+ e^{2 b(t)} {\rm e}^{34}+ e^{2 c(t)} {\rm 
e}^{56}\,,
\end{equation}
with $a(t),b(t),c(t)$ being arbitrary for the moment. 
Since the general solution of (\ref{flow}) for $\Psi$ takes the form
\begin{equation}
\Psi(t)=e^{\int_{t_0}^t Q(t') dt'} \widehat{\Psi}
\end{equation}
and $\widehat{\Psi} \wedge 
\overline{\widehat{\Psi}} = -\frac{4 i}
{3} \widehat{J}\wedge\widehat{J}\wedge\widehat{J}$, we obtain a 
relation between the singlet component of the flux and the size 
deformations of the metric
\begin{equation}
\label{eq:flux_fun}
Q(t)=\dot{a}(t)+\dot{b}(t)+\dot{c}(t) \,.
\end{equation}
In this way the conditions on the base (\ref{specialhermitian}) and 
the second equation in (\ref{flow}) are satisfied by construction.
In order to fulfill the first flow equation (\ref{flow}), we can fix 
the $A$ component of the flux as 
\begin{equation}
A=\frac{1}{3} e^{2 a(t)} (-2 \dot{a}+\dot{b}+\dot{c}) {\rm e}^{12}+
\frac{1}{3} e^{2 b(t)} (-2 \dot{b}+\dot{c}+\dot{a}) {\rm e}^{34}+
\frac{1}{3} e^{2 c(t)} (-2 \dot{c}+\dot{a}+\dot{b}) {\rm e}^{56}\,.
\end{equation}
Furthermore $S$ can be computed from (\ref{specialhermitian}):
\begin{equation}
S=-\frac{1}{2}e^{2c} \left({\rm e}^{136}-{\rm e}^{246}-{\rm e}^{145}-{\rm
    e}^{235}\right)\,.
\end{equation}

In order to have a full solution we need to solve the Bianchi identity for $G$.
This reads
\begin{eqnarray}
dG&=&\frac{1}{3} e^{2a+2b} \left(-2 \ddot{a}-2\ddot{b}+ \ddot{c}-4
\dot{a}^2 -4 \dot{b}^2 +2 \dot{b}\dot{c}+2\dot{a}(-4 \dot{b}+\dot{c})\right)
{\rm e}^{1234}\wedge v  \nonumber \\
&+& \frac{1}{3} e^{2b+2c} 
\left(-2 \ddot{b}-2\ddot{c}+ \ddot{a}-4
\dot{b}^2 -4 \dot{c}^2 +2 \dot{c}\dot{a}+2\dot{b}(-4 \dot{c}+\dot{a})\right)
{\rm e}^{3456}\wedge v  \label{eq:dg_iwa}\\
&+& \frac{1}{3} e^{2c+2a} 
\left(-2 \ddot{c}-2\ddot{a}+ \ddot{b}-4
\dot{c}^2 -4 \dot{a}^2 +2 \dot{a}\dot{b}+2\dot{c}(-4 \dot{a}+\dot{b})\right)
{\rm e}^{5612}\wedge v -2 e^{2c} {\rm e}^{1234} \wedge v \nonumber ,
\end{eqnarray}
where we have used (\ref{eq:flux_fun}). The magnetic 6-form current is
\begin{equation}
{\cal J}_6 = e^{4 \Delta} dx^0 \wedge dx^1 \wedge dx^2 \wedge dx^3 \wedge
\left(\rho_{12}(t) {\rm e}^{12} + \rho_{34}(t) {\rm e}^{34}+ 
\rho_{56}(t) {\rm e}^{56}\right)
\end{equation}
where
\begin{eqnarray}
\rho_{12}(t) &=& \frac{1}{3} e^{2b+2c} 
\left(-2 \ddot{b}-2\ddot{c}+ \ddot{a}-4
\dot{b}^2 -4 \dot{c}^2 +2 \dot{c}\dot{a}+2\dot{b}(-4 \dot{c}+\dot{a})\right) 
\nonumber\\
\rho_{34}(t) &=& \frac{1}{3} e^{2c+2a} 
\left(-2 \ddot{c}-2\ddot{a}+ \ddot{b}-4
\dot{c}^2 -4 \dot{a}^2 +2 \dot{a}\dot{b}+2\dot{c}(-4 \dot{a}+\dot{b})\right)\\
\rho_{56}(t) &=& \frac{1}{3} e^{2a+2b} \left(-2 \ddot{a}-2\ddot{b}+ \ddot{c}-4
\dot{a}^2 -4 \dot{b}^2 +2 \dot{b}\dot{c}+2\dot{a}(-4 \dot{b}+\dot{c})\right)
-2 e^{2c}\,.\nonumber
\end{eqnarray}

One can first look for solutions with no sources, i.e. $dG=0$ or equivalently
$\rho_{12}(t)=\rho_{34}(t)=\rho_{56}(t)=0$. Analyzing however the resulting
system of differential equations seems quite difficult in practice and
we postpone this for future work. Hence, we are left with the possibility of
interpreting the non-zero piece of $dG$ as due to M5--branes wrapping 2--cycles
in the internal 7-dimensional manifold. Since this manifold is 
a fibration of the Iwasawa manifold over an interval, we can actually
imagine that our solution corresponds to a fibration of a configuration
of wrapped M5--branes on calibrated 2-cycles of the Iwasawa. 

Since $\widehat{J}$ is a generalized calibration on the Iwasawa, we conclude
that branes wrapping the holomorphic 2-cycles $u=v=0$ and $z=u=0$
are BPS.  We can obtain such an
interpretation of our example if we demand  $\rho_{56}(t) =0$. 
A simple way to solve this equation is to set $a=b=0$. 
Then $c(t)=- 
{\rm log}\left(\sqrt{6}t\right)$ 
and $\rho_{12}(t) = \rho_{34}(t) = -\frac{1}{3 t^4}$ 
give the density of wrapped M5--branes in the transverse space. 
Notice that the 5-branes are distributed uniformly on the Iwasawa manifold
and the only non-trivial part of the solution comes from the fibration
over $I$. Finally, to specify completely the solution we present the 11-dimensional metric:
\begin{equation}
ds^2_{11}(x,t)= t^{2/3} \eta_{\mu \nu} dx^{\mu} dx^{\nu}+
\left(dz d{\overline z}+ dv d{\overline v} + \frac{1}{6t^2} (-du+z dv)  
(-d{\overline u}+{\overline z} d{\overline v})\right)
+ dt^2 .
\end{equation}

Notice that since all holomorphic 2--cycles in a complex manifold are 
calibrated, one can seek more solutions. 
However, our purpose here was to illustrate
the general technique and to discuss a simple example.

\subsubsection{Generalization of Strong K\"ahler Torsion (SKT) manifolds}

For the case of zero flux in the singlet component of $G$, the special 
hermitian manifolds one has to use as base of the full 7--dimensional 
spaces have to satisfy an interesting relation.
Since $Q=0$, from (\ref{specialhermitian}) and (\ref{flow}) it follows that $\widehat{d} A=\dot{S}$. 
Using the flow equations and the fact that $J \wedge S =0$ we find that
the source-free Bianchi identity becomes
\begin{equation}
\widehat{d} * \widehat{d} {J} = \frac{1}{2} \frac{d^2}{dt^2} 
({J}\wedge{J}) \wedge v\,.
\end{equation}

A simple way to solve this equation is to assume
a $t$--dependence of the form $J(t)=e^{\frac{m}{2} t} \widehat{J}$. Then we
get an equation for the 2--form $\widehat{J}$ that reads
\begin{equation}
(\widehat{d}^{\dagger}\widehat{d} \widehat{J}-m^2) \widehat{J} = 0,
\end{equation}
where we have used $* \left({J}\wedge{J}\wedge v\right)=-2 {J}$. 
Being complex, this manifold has to satisfy a generalization of the 
SKT condition $\partial \overline \partial \widehat J =0$.
It would be interesting to find explicit examples of 
6--dimensional manifolds with almost complex
structures satisfying this equation.

\subsection{Case II: $q\neq 0$}

Let us now generalize the previous discussion for
non--zero $\phi=\phi(y,t)$. 
We will show that the Fayyazuddin-Smith
solution \cite{Fayyazuddin:1999zu,Brinne:2000nf,Smith:2002wn,Husain:2003df} 
falls in this class.

We assume an ansatz for
the 7--dimensional metric of the following form
\begin{equation}
ds^2_{7}(y,t)= e^{p \phi} ds^2_{6}(y,t) + e^{2 \phi} dt^2\,.
\end{equation}
{From} (\ref{eqdDelta}) and the 
condition (\ref{eqdv}) on $v = e^{\phi} \,dt$ we can specify the warp factor 
dependence on the coordinates of the 6--dimensional base $y$ and the fiber 
$t$
\begin{equation}
\sigma\equiv \widehat{d}\Delta=-\frac{1}{2}\widehat{d}\phi,\;\;\;\;\;
\dot{\Delta}=-\frac{1}{3} Q e^{\phi} \,.
\end{equation}
This dependence is further restricted by the consistency condition 
(\ref{constr0}) which in the case at hand becomes
\begin{equation}
\widehat{d}Q=- Q \,\widehat{d} \phi\,.
\end{equation} 
{From} this latter we can see that 
$\widehat{d} \dot{\Delta}=\widehat{d}\dot{\phi}=0$  and that therefore 
the warp factor splits as $\Delta(x,t)=\Delta_1(x)+\Delta_2(t)$.

Now, denoting with a hat the 2-- and 3--forms of the $SU(3)$ structure 
on the 6--dimensional base
\begin{equation}
J=e^{p\phi} \,\widehat{J}, \;\;\;\;\;
\Psi=e^{\frac{3}{2} p \phi}\, \widehat{\Psi} \,,
\end{equation}
by using (\ref{eqdv2})--(\ref{eqdpsi2}) we obtain once again conditions 
on the geometry of the base space
\begin{equation}
\left\{   
\begin{array}{rcl}
\widehat{d}\widehat{J} &=& -2\, e^{-\frac{1}{2}\phi} \,S \\[2mm]
\widehat{d}\widehat{\Psi}&=&\displaystyle\frac{3}{4}\, \widehat{d}\phi \wedge \widehat{\Psi}
\end{array}
\right.
\label{balanced}
\end{equation}
and the flow equations describing the uplift to seven dimensions
\begin{equation}
\left\{   
\begin{array}{rcl}
\displaystyle
\frac{\partial \widehat{J}}{\partial t} &=&\displaystyle \frac{2}{3} 
e^{\phi} Q\, \widehat{J} - 2
e^{\frac{\phi}{2}}\, A -\frac{1}{2}\, \dot{\phi}\, \widehat{J}\\[2mm]
\displaystyle
\frac{\partial \widehat{\Psi}}{\partial t} &=&\displaystyle Q 
e^{\phi}\, \widehat{\Psi} -\frac{3}{4}\,
 \dot{\phi} \,\widehat{\Psi} \,.
\end{array}
\right.
\label{flow_two}
\end{equation}
The conditions (\ref{balanced}) on the $SU(3)$ structure 
mean that we deal with the so-called balanced manifolds, which are complex
but not K\"ahler. 
Notice that we have chosen $p=1/2$ in order to make 
the equation for $\widehat{d}\widehat{J}$ simpler. If we had kept an arbitrary
$p$, we would have a non-zero ${\cal W}_4$ class and the manifold
would be conformally balanced. This is due to the fact that 
under conformal transformations of the 6-dimensional metric,
the combination $3 {\cal W}_4 + 2 {\cal W}_5$ remains constant.

It is straightforward now to verify that 
the Fayyazudin-Smith solution, as described for example in
\cite{Brinne:2000nf}, falls in the above class 
by identifying the metric and warp factors as
\begin{eqnarray}
H_1 = e^{\Delta}\,, \quad
H_2 = e^{\phi}\,, \quad
2 G_{M \overline{N}} dz^M dz^{\overline N} = e^{\frac{\phi}{2}}  ds^2_{6} \,.
\end{eqnarray}
An interesting consistency condition follows from the fact 
that $\widehat{J}^3 = \sqrt{g_6} d^6 y$, where $g_6(y,t)$ the 
metric tensor corresponding to $ ds^2_{6}(y,t)$.
This condition, which reads
\begin{equation}
\frac{\partial}{\partial t} {\rm log}\sqrt{g_6} = -6 \left(\frac{1}{4} \dot{\phi}
+ \dot{\Delta}\right)\,, 
\end{equation}
is useful in establishing the relation of the above solution 
to \cite{Fayyazuddin:1999zu,Brinne:2000nf,Husain:2003df}.

\section{Type IIA reduction}

We now turn to the discussion of 7--manifolds with isometries, so that 
we can reduce the solution to type IIA strings.
As we already saw in the introduction, one has to distinguish two 
cases according to the relation between the vector describing the isometry 
of the solution and the vector $v$ defining the $SU(3)$ structure in 7 
dimensions. 
First we analyze the case of $v$ being proportional to the isometry and 
then discuss the more complicated situation arising from a reduction 
in a direction orthogonal to $v$.

\subsection{Case I: reduction to type IIA along $v$}
\label{redv}

When the 7--manifold has an isometry, we can find a set of coordinates 
so that this isometry is described by a constant Killing vector 
field $\partial/\partial z$.
The $SU(3)$ structure in 7 dimensions is described also by a globally 
defined vector field $v$ and we are now going to discuss the 
possibility of such a vector being proportional to the Killing vector
$\partial/\partial z$. 
Since we need $v$ to be globally defined,
the dual one--form must be defined as
\begin{equation}
v = e^{\beta \phi}\, dz\,,
\label{eqvglob}
\end{equation}
$dz$ being the differential associated to the isometry.
As opposed to a generic IIA reduction of a 7--manifold with 
isometries, one cannot introduce a 
non--trivial gauge potential in  (\ref{eqvglob}) and hence 
the RR 2-form flux in type IIA is zero: $F=0$.

The fact that $v$ is globally defined means that we have an almost
product structure and the metric of ${\cal M}_7$ takes the form
\begin{equation}
ds^2_{7}(y)= e^{-2 \alpha \phi(y)} d\tilde s^2_{6}(y) + 
v (y) \otimes v (y)\,,
\label{metricv}
\end{equation}
where there is no $z$--dependence since the corresponding vector field
is Killing and the $\alpha$ and $\beta$ parameters have been introduced 
in order to maintain the freedom of choice of the 10--dimensional frame.
The string frame is obtained by setting $\beta = 2\alpha = 2/3$. 
{From} (\ref{metricv}), one can see that the 7--dimensional $SU(3)$ structure 
naturally induces an $SU(3)$ structure in 6 dimensions
\begin{equation}
\widetilde{J} =e^{2 \alpha\phi}\,J\,,\quad
\widetilde{\Psi} = e^{3 \alpha\phi} \,\Psi\,.
\label{6dstr}
\end{equation}
This implies that we can discuss the resulting type IIA 
compactifications in the presence of fluxes in terms of such 
structures and their intrinsic torsion classes.

The computation of the $SU(3)$--torsion classes for the 6--dimensional 
manifolds follows from the application of the above definitions 
(\ref{eqvglob}), (\ref{metricv}) and (\ref{6dstr}) to the 
conditions (\ref{eqdv2})--(\ref{eqdpsi2}).
The equation on $dv$ will not reduce to constraints on the 6--dimensional 
torsion, but it will further constrain the possible solution.
{From} the explicit computation $dv  = \beta d\phi \wedge v $ we conclude that
\begin{equation}
\sigma=-\frac{\beta}{2} d\phi\,.
\end{equation}
It can also be noted that even though we set to zero the gauge field from 
first principles, its introduction would have resulted again in finding $F 
=0$ as the analysis of $dv$ shows.
Other constraints on the fluxes as well as the definition of the 
6--dimensional torsion can be obtained now by the evaluation of $dJ$ 
and $d\Psi$ and their comparison with (\ref{eqdJ2})--(\ref{eqdpsi2}).
The $SU(3)$ singlet and adjoint components of the 11--dimensional flux
are vanishing, i.e. $A=0$ and  $Q=0$. 
The latter implies that we can identify the warp factor with the 
10--dimensional dilaton $\Delta=-\frac{\beta}{2} \phi$. 
The differentials of the 6--dimensional forms read
\begin{eqnarray}
d\widetilde{J}&=&-2 e^{2\alpha\phi} S + 
\left(2\alpha+\frac{\beta}{2}\right) d\phi \wedge
\widetilde{J}\\
d\widetilde{\Psi}&=&3\left(\alpha+\frac{\beta}{2}\right) d\phi \wedge
\widetilde{\Psi} .
\end{eqnarray}
{From} these we can read eventually the relevant intrinsic torsion classes 
\begin{equation}
{\cal W}_1 =0,\;\;{\cal W}_2=0,\;\;{\cal W}_3=-2 e^{2\alpha\phi} S, \;\;
{\cal W}_4= \left(2\alpha+\frac{\beta}{2}\right) d\phi ,\;\; {\cal 
W}_5=-3\left(\alpha+\frac{\beta}{2}\right)
d\phi .
\end{equation}

Let us now comment on the solution.
The original configuration in M--theory involved a warped solution 
starting from a 4--form flux given by 
\begin{equation}
G=v  \wedge (J\wedge W + S)\,.
\label{eqfluxvred}
\end{equation}
The reduction to type IIA along $v$ gives generically a complex 
manifold with torsion.
In the string frame, where $2\alpha=\beta$, the torsion components 
satisfy a constraint given by $2 {\cal W}_4+ {\cal W}_5=0$, with both ${\cal 
W}_4$ and ${\cal W}_5$ being exact and proportional to the exterior 
derivative of the dilaton.
${\cal W}_3$ remains free and related to the flux. 
This result is precisely the same as that found in 
\cite{Cardoso:2002hd,Gauntlett:2003cy} where the ``common sector'' of 
type I/II and heterotic theory was analyzed.
Indeed (\ref{eqfluxvred}) shows that the only flux present in the 
reduced theory is the NS--NS 3--form flux. 
Moreover, one can simply realize that $H^{(3,0)}=H^{(0,3)}=0$ because 
of the constraint (\ref{eqGcond}).

If we assume  
that $v$ itself is Killing, i.e. $\beta=0$, then the resulting 
6--manifold is special--hermitian.
The solution will not show a warp factor and the only torsion class 
different from zero will be ${\cal W}_3$, originating from
the primitive part of the flux.

\

\subsection{Case II: reduction to type IIA along  $\tau \perp v$}

Under the general assumption that the 7--dimensional
manifold ${\cal M}_7$ used in the solution of M--theory with 4--form 
flux has a Killing isometry, we have a metric ansatz of the form
\begin{equation}
ds^2_{7}(y)= e^{-2 \alpha \phi(y)} d\tilde s^2_{6}(y) + 
\tau(y) \otimes \tau(y)
\label{metrIIA}
\end{equation}
where $\tau(y)=e^{2 \beta \phi(y)} (dz + A(y))$ is a 
1--form describing a non--trivial $U(1)$ fibration over the 6--dimensional
manifold ${\cal M}_6$ parameterized by $x$.
We will now discuss the case that $\tau$ is not globally defined, so 
that one can assume $\tau \perp v$. 
Performing the reduction to type IIA in this way implies that $v$ is 
inherited by the 6--dimensional manifold so that its group structure 
is at least reduced to $SO(5)$.
On the other hand, we know that the requirement of preserving 
some supersymmetry in type IIA imposes the existence of an $SU(3)$ 
structure for the internal manifold. Hence, we finally expect
a 6--dimensional $SU(2)$ structure. 

Since there is a $U(1)$-worth of $SU(2)$ embeddings in $SU(3)$, one 
can reconstruct the full $SU(3)$ in different ways.
Obviously, this degeneracy does not change the physics, and therefore we 
arbitrarily fix the extra phase and define
\begin{eqnarray}
v  &=& e^{-\alpha \phi} \,\widetilde{v}\,, \\
J &=& e^{-2\alpha \phi}\, \widetilde{J} + e^{-\alpha \phi}\, \widetilde{w} 
\wedge \tau\,, \label{Jddef}\\
\Psi &=&  e^{-2\alpha \phi} \widetilde{K} \wedge \left( e^{-\alpha 
\phi}\, \widetilde{w} + i \, \tau \right) \,.
\end{eqnarray}
Here the 6--dimensional forms $\widetilde{v}, \widetilde{w}, \widetilde{J}, \widetilde{K}$
characterize the $SU(2)$ structure (the complex 1--form of subsection 1.1
is given by $w  = \widetilde{v}+i\, \widetilde{w}$).
We also have 
\begin{equation}
d\tau = \beta \,d\phi \wedge \tau + e^{\beta \phi} \,F\,, 
\end{equation}
where $F$ is the 2--form field strength of the type IIA gauge field.

Besides the metric assumption (\ref{metrIIA}) and the definition of 
the $SU(2)$ structure given above, we do not impose extra constraints 
on the fluxes for the moment.
The usual computation of the intrinsic torsion leads to the 6--dimensional $SU(2)$
and $SU(3)$ structures coming from the supersymmetry constraints on 
$dv , dJ$ and $d\Psi$. 
After some tedious but straightforward computations, one obtains the 
conditions on the torsion classes 
\begin{eqnarray}
\label{vsu2}
d\widetilde{v} &=& (\alpha d\phi - 2 \sigma) \wedge \widetilde{v}, 
\\[2mm]
d\widetilde{w} &=& \Big( (\alpha-\beta) d\phi - \sigma + \frac{2}{3} Q
  e^{-\alpha\phi} \widetilde{v }\Big)\wedge \widetilde{w} - 2 e^{\alpha 
  \phi} \,(\tau \,\lrcorner\; S)
+ 2 \widetilde{v }\wedge (\tau \,\lrcorner\; A), \\[2mm]
d\widetilde{J} &=& e^{(\alpha+\beta)\phi} F \wedge \widetilde{w} + \Big(2\alpha d\phi
-\sigma + \frac{2}{3} Q e^{-\alpha\phi} \widetilde{v }\Big)\wedge \widetilde{J} - 2 e^{2\alpha\phi}
S|_{h} -2 e^{\alpha\phi}\, \widetilde{v } \wedge A|_h ,\\[2mm]
d\widetilde{K} &=& \Big((2\alpha-\beta) d\phi - 3 \sigma + Q\, 
\widetilde{v} \, e^{-\alpha\phi} \Big)\wedge \widetilde{K} .
\label{Ksu2}
\end{eqnarray}
In the above formulas $S|_{h}$ and $A|_{h}$ mean the horizontal 
part of $S$ and $A$ with respect to $\tau$.
In addition the following compatibility constraint arises 
\begin{equation}
\label{eq:cons1}
(-2 \beta d\phi -\sigma + \frac{2}{3} Q \widetilde{v } e^{-\alpha \phi}) \wedge
\widetilde{K}\wedge\widetilde{w} + i e^{(\alpha+\beta)\phi} \widetilde{K}\wedge F -
2 e^{\alpha\phi} \widetilde{K}\wedge (\tau\,\lrcorner\; S) + 2 \widetilde{K}\wedge\widetilde{v }\wedge
(\tau \,\lrcorner\; A) = 0 \,.
\end{equation}
Though this seems quite a cumbersome expression, we will see 
later how it can be used to extract 
useful physical information.

\subsubsection{10--dimensional vacua with 2--form flux}

A purely geometrical solution of M--theory with at least one 
isometry can be reduced to a type IIA background with non--trivial 
dilaton and 2--form flux.
Though these solutions have been already analyzed in terms of 
$SU(3)$--structures in \cite{chiossi,Kaste:2002xs,Kaste:2003dh}, 
we will analyze them once again in terms of $SU(2)$ structures 
in order to fit them in our framework.

Once all the 4--form flux components are turned off, 
the intrinsic torsions of the 6--dimensional $SU(2)$ structure read
\begin{equation}
\label{SU22flux}
\begin{array}{rcl}
d\widetilde{v} &=& \alpha \,d\phi \wedge \widetilde{v}\,, \\[2mm]
d\widetilde{w} &=& (\alpha-\beta) \,d\phi \wedge \widetilde w\,, \\[2mm]
d\widetilde{J} &=& e^{(\alpha+\beta)\phi}\, F \wedge \widetilde{w} + 2\alpha 
\,d\phi
\wedge \widetilde{J}\,, \\[2mm]
d\widetilde{K} &=& (2\alpha-\beta)\, d\phi \wedge \widetilde{K} \,
\end{array}
\end{equation}
and the consistency constraint (\ref{eq:cons1}) simplifies to
\begin{equation}
\label{conF}
2\beta \, d\phi\wedge\widetilde{K}\wedge\widetilde{w}={\rm i}  \,
e^{(\alpha+\beta)\phi} \,\widetilde{K}\wedge F \,.
\end{equation}

In order to compare the above expressions with the known results given 
in terms of $SU(3)$ structures, we have to introduce an almost complex 
structure and the associated (3,0)--form in 6--dimensions.
In doing so one has to face a $U(1)$ ambiguity following from the 
embedding of $SU(2)Ê\subset SU(3)$.
Since such ambiguity should not result in a physical difference, we 
will simply fix it in a convenient way and choose 
\begin{eqnarray}
\label{eq:defJ}
J&=&J^2+\widetilde{w}\wedge\widetilde{v} \,,\\
\label{eq:defOm}
\Omega& =& (J^3 + {\rm i} J^1)\wedge(\widetilde{w}+{\rm i}\, \widetilde{v}) \,,
\end{eqnarray}
where $\widetilde{J}=J^1$ and  $\widetilde{K}=J^2+{\rm i}\,J^3$. 
Some of the torsion classes follow now straightforwardly from $dJ$: 
\begin{equation}
{\cal W}_1=0, \;\;{\cal W}_3=0, \;\;{\cal W}_4=(2\alpha-\beta) \,d\phi.
\end{equation}

Before proceeding with the rest of the torsion classes, we note that
the constraint (\ref{conF}) implies  $\widetilde{v } \,\lrcorner\; d\phi =0$
and $\widetilde{v } \,\lrcorner\; F =0$.
This, together with (\ref{eq:defJ}), constrains the possible  
2--form field strength, whose general form is given by 
\begin{equation}
\label{eq:solutionF}
F = 2 \beta\, e^{-(\alpha+\beta)\phi}\, \left(
-\frac{1}{2}\,  (\widetilde{w}\,\lrcorner\; d\phi) \, J^1 -(d\phi\,\lrcorner\; J^1) \wedge
\widetilde{w}\right) + F_{0} \,.
\end{equation}
Here $F_{0}$ denotes the primitive part of $F$ with respect to the three
4--dimensional almost complex structures, i.e. 
$F_{0} \wedge J^i =0$ for $i=1,2,3$.
It should be noted that $F$
is primitive with respect to the $SU(3)$ structure defined by 
(\ref{eq:defJ})--(\ref{eq:defOm}), i.e.
$ J \,\lrcorner\; F = 0$ or equivalently $J \wedge J \wedge F = 0$. 
From this expression one can also obtain 
\begin{equation}
\label{monop}
F \, \lrcorner\; {\rm Im}\Omega = -2 \, \beta\, 
e^{-(\alpha+\beta)\phi}\, d\phi\,,
\end{equation}
which is the generalized monopole equation noticed in \cite{Kaste:2003dh}.

The remaining torsion classes are now determined by computing $d\Omega$: 
\begin{equation}
\label{W2twoflux}
{\cal W}_2 = - e^{(\alpha+\beta)} \,F - \beta\, (d\phi \,\lrcorner\; {\rm  Im}\Omega),\;\;
{\cal W}_5=-(3\alpha-\beta)\,d\phi \,.
\end{equation}
One can verify that the expression for ${\cal W}_2$ reduces to
the primitive $(1,1)$ piece of the 2-form flux with respect to $J$ 
using (\ref{eq:solutionF}).
In order to do so one first writes $F$ as a sum of irreducible components
\begin{equation}
F= m\, J +  F_{0}^{(1,1)} + (F^{(2,0)}+F^{(0,2)}) \,.
\end{equation}
{From} the general solution (\ref{eq:solutionF}) one gets $m=0$
and $(F^{(2,0)}+F^{(0,2)}) = \frac{1}{2}\, (F\,\lrcorner\; \Omega_-) \,\lrcorner\; \Omega_-=
- \beta e^{-(\alpha+\beta)\phi} \,(d\phi \,\lrcorner\; \Omega_- ).$ 
Finally, by using this in (\ref{W2twoflux}), one gets the expected 
result: ${\cal W}_2 = - e^{(\alpha+\beta)\phi} F_{0}^{(1,1)}$.

A background satisfying the above relations can be constructed from 
a deformation of ${\mathbb T}^4 \times {\mathbb R}^2$ by functions of the dilaton
\begin{equation}
d\tilde s_6^2 =  \sum_{i=1}^4 dy_i^2 + e^{-2\alpha\phi} 
\,dy_5^2 + e^{2\alpha\phi} \,dy_6^2\,,
\label{eqmetr}
\end{equation}
where we assume that we are in the string frame, i.e. $2\alpha = \beta$.
This metric satisfies (\ref{SU22flux}) with the definitions
\begin{equation}
\begin{array}{rclrcl}
\widetilde v & = & e^{\alpha\phi}\,dy_{6}\,,  &
\widetilde w & = & e^{-\alpha\phi}\, dy_5\,,  \\[2mm]
\widetilde J & = & dy_1 \wedge dy_2 + dy_3 \wedge dy_4\,,  &
F &=& 2 \alpha \,\widetilde J\,,
\end{array}
\label{eqsoluz}
\end{equation}
and by choosing the dilaton to be a logarithmic function of $y_5$.
A simple extension is given by adding more warpings in the metric 
depending on the dilaton
\begin{equation}
d\tilde s_6^2 = e^{-2\alpha\phi}\,\left(dy_3^2+ dy_4^2+ 
dy_5^2\right) + e^{2\alpha\phi} \,\left(dy_1^2 + dy_2^2+dy_6^2\right)\,.
\label{eqwarped}
\end{equation}
In this case the flux is not given simply in terms of $\widetilde J$, 
but we get 
\begin{equation}
\begin{array}{rclrcl}
\widetilde J & = & e^{2\alpha\phi}\,dy_1 \wedge dy_2 + e^{-2\alpha\phi}dy_3 \wedge dy_4\,,  &
F &=& 4 \alpha \, dy_3 \wedge dy_4 \,.
\end{array}
\label{eq:soluz}
\end{equation}
One can check that the full 10--dimensional IIA metric corresponding 
to (\ref{eq:soluz}) has 7--dimensional Poincar\'e invariance and therefore 
it can be interpreted as a configuration of smeared D6--branes.

A twisted version of the solution just presented is shown in \cite{Kachru:2002sk}.
This is obtained by T-dualizing three times a type IIB solution given 
by ${\mathbb T}^6/{\mathbb Z}_2$ with 3--form fluxes. 
The type IIA metric in the string frame is
\begin{equation}
d\tilde s_6^2 = e^{-2\phi/3}\,\left[\left(dx_1 + 2 x_2 \, dx_3\right)^2 +  
\left(dy_1 + 2 y_2 \, dx_3\right)^2+ dy_3^2\right]
+ e^{2\phi/3}\left[dx^2_2 + dx_3^2 + dy_2^2\right]\,,
\label{eqtori2twist}
\end{equation}
and there is a non--trivial 2--form flux 
\begin{equation}
F = 2 \left(dx_1 + 2 x_2 \, dx_3\right)\wedge dy_2 + 2\left(dy_1 + 2 
y_2 \, dx_3\right) \wedge dx_2\,.
\label{eqfluxtwist}
\end{equation}
In the presentation of \cite{Kachru:2002sk}, the dilaton was neglected, 
though it should be a non--trivial function of $x_2$, $x_3$ and $y_2$.
This means that one cannot fit this solution in the scheme presented above, unless 
the dilaton is determined.
Moreover, neglecting the dilaton implies that the 5--form flux of the 
original type IIB solution was neglected and, accordingly, 
all of its contributions in the dualization procedure.

\subsubsection{10--dimensional vacua with 2--form and 3--form flux}

Before dealing with examples of the most general configuration of 
fluxes, we want to show solutions of type IIA string theory in the 
presence of 2--form and 3--form flux.
{From} the 11--dimensional point of view this means that the 4--form flux $G$
satisfies $G \wedge \tau=0$. 
Using the general expression for $G$ we conclude that the $SU(3)$ singlet 
and adjoint components are vanishing and that the vector and double symmetric 
tensor are not arbitrary. Instead 
\begin{equation}
Q=0\,, \;\; A=0\,, \;\; \sigma=\lambda\, \widetilde{w}\,, \;\; S=\widetilde{w}\wedge 
X\,,
\end{equation} 
where $X$ should be a primitive 2--form with respect to $J^i, i=1,2,3$.

We can analyze this case as an extension of the previous one, by 
describing the 6--dimensional intrinsic torsions in terms of the ones  
in (\ref{SU22flux}) incorporating of course the new flux contributions.
In this way one gets
\begin{equation}
\begin{array}{rcl}
d\widetilde{v} &=& d\widetilde{v}_{old} - 2 \lambda \,\widetilde{w} 
\wedge \widetilde{v}\,, \\[2mm]
d\widetilde{w} &=& d\widetilde{w}_{old}\,, \\[2mm]
d\widetilde{J} &=& d\widetilde{J}_{old} - (\lambda \,\widetilde{w} \wedge \widetilde{J} + 
2 e^{2\alpha\phi}\,
X \wedge \widetilde{w})\,, \\[2mm]
d\widetilde{K} &=& d\widetilde{K}_{old} - 3 \lambda \,\widetilde{w} \wedge 
\widetilde{K}\,,
\end{array}
\end{equation}
where $d\widetilde{v}_{old}, d\widetilde{w}_{old}, d\widetilde{J}_{old}, 
d\widetilde{K}_{old}$ are the differentials in  (\ref{SU22flux}).
We note now that even though the constraint  (\ref{eq:cons1})  remains 
the same as (\ref{conF}), the precise
form of the solution for the 2--form flux changes because $dJ^1$ and 
$dJ^2$ are different. 
The new solution is
\begin{equation}
\label{newF}
F = 2 \beta e^{-(\alpha+\beta)\phi} \,\left(
-\frac{1}{2}\,  (\widetilde{w}\,\lrcorner\; d\phi) \, J^1 -(d\phi\,\lrcorner\; J^1) \wedge
\widetilde{w}\right) -2 \lambda  \,e^{-(\alpha+\beta)\phi}  \,J^1 + 
F_{0} \,.
\end{equation}
Notice the appearance of the extra term proportional to $\lambda$ 
which changes the generalized monopole equation (\ref{monop}) to
\begin{equation}
F \, \lrcorner\; {\rm Im}\Omega = -2 \, \beta\, 
e^{-(\alpha+\beta)\phi}\, d\phi - 4 \,e^{-(\alpha+\beta)\phi}\, 
\lambda \, \widetilde w\,.
\end{equation}

As we did in the previous case, we will now interpret the above 
constraints in terms of $SU(3)$ structures determined by a $J$ and $\Omega$
defined as in (\ref{eq:defJ})--(\ref{eq:defOm}). 
Although this will not change the mathematical description we just 
presented, it will help us improve our intuition about the possible 
solutions.
The torsion conditions on the almost complex structure are obviously 
the same as before, except for the addition of an extra piece depending on the 
3--form flux:
\begin{equation}
dJ=dJ_{old} - 3 \,\lambda \,J^2\wedge\widetilde{w}=dJ_{old} + J \wedge 
(-3 \lambda\, \widetilde{w})\,.
\end{equation}
Hence, the torsion classes determined by $J$ are now
\begin{equation}
{\cal W}_1=0, \;\;{\cal W}_3=0, \;\;{\cal W}_4=(2\alpha-\beta) d\phi - 
3 \lambda \widetilde{w} \,.
\end{equation}
The same applies to $d\Omega$, which reads
\begin{equation}
d\Omega=d\Omega_{old} -5 {\rm i}\, \lambda \,
J^3\wedge\widetilde{w}\wedge\widetilde{v} +
3 \lambda\,  J^1\wedge\widetilde{w}\wedge\widetilde{v} + 2 e^{2\alpha\phi} 
\,X  \wedge\widetilde{w}\wedge\widetilde{v} \,.
\end{equation}
The remaining torsion classes are more easily understood if one puts 
the extra terms in irreducible form 
\begin{equation}
d\Omega=d\Omega_{old} + \Omega \wedge \frac{5}{2} \,\lambda \,
(\widetilde{w} - {\rm i}\,\widetilde{v})
+ 2\, J \wedge (e^{2\alpha\phi} \,X - \lambda\, J^1) \,.
\end{equation}
{From} this, one concludes that
\begin{equation}
\begin{array}{rcl}
{\cal W}_2 &=& \left(- e^{(\alpha+\beta)\phi}\, F - \beta \,(d\phi \,\lrcorner\; \Omega_- ) - 2 
\lambda\, J^1\right) + 2 e^{\alpha\phi} \,X, \\[2mm]
{\cal W}_5&=&-(3\alpha-\beta)\,d\phi + 5 \lambda\, \widetilde{w} \,.
\end{array}
\end{equation}
The ${\cal W}_2$ class contains again an extra piece depending on 
$\lambda$, but inspection of (\ref{newF}) shows that this is
exactly canceled by the $\lambda$--dependent term that appears there.
This means that once again we are left with the primitive $(1,1)$
piece of $F$. 
Finally, we can describe this class explicitly in terms of primitive 
$(1,1)$ forms
\begin{equation}
{\cal W}_2 =  -F_{0}^{(1,1)} +  2 e^{\alpha\phi} \,X \,.
\end{equation}


So far the torsion classes given above show that the manifolds one has 
to use in order to find solutions of type IIA strings in the presence 
of fluxes have to be non--complex.
However, it should be noted that for certain choices of the 2--form flux 
one can obtain an integrable complex structure. 
Indeed, by choosing the fluxes so that $F_{0}^{(1,1)} =   2 
e^{\alpha\phi}\, X$, the classes ${\cal W}_1$ and ${\cal W}_2$ are vanishing and 
therefore the resulting 6--dimensional manifold is complex.
Even more interesting is that by selecting $\beta \,d\phi =  \lambda\,
\widetilde{w}$ one can satisfy the extra condition 
\begin{equation}
3\, {\cal W}_4 + 2\, {\cal W}_5 =0\,.
\end{equation}
This condition describes manifolds which are conformal rescalings of
Calabi--Yau spaces \cite{chiossi}. 
Hence, for these particular combinations of 2-- and 3--form
fluxes we can obtain relatively simple type IIA solutions with full metric 
in the string frame
\begin{equation}
ds^2_{10} = e^{2 \phi}\, \eta_{\mu \nu} dx^{\mu} dx^{\nu}+ 
e^{-2\phi}\, ds^2_{CY}(y) \,,
\end{equation}
where $ds^2_{CY}(y)$ is a 6--dimensional CY metric. 
One should be careful however to check that the corresponding
$SU(2)$ structure has also the required intrinsic torsion and
that the Bianchi identity for the 3--form flux is satisfied. 
In these cases, the two--form flux satisfies again a generalized monopole 
equation like (\ref{monop}), but with a different coefficient
\begin{equation}
F \, \lrcorner\; {\rm Im}\Omega = -4  \, 
e^{- \phi}\, d\phi \,.
\end{equation}

\subsubsection{10--dimensional vacua with general fluxes}

In the previous sections we specified the general results of
(\ref{vsu2})--(\ref{Ksu2}) for solutions with only 2--form flux $F \neq
0$ or for solutions with both 2--form and 3--form fluxes $F \neq 0$, $H
\neq 0$.
We are going now to present a way to construct solutions where all the
10--dimensional fluxes are turned on.

Our strategy
is partially inspired by \cite{Goldstein:2002pg}.
The starting point is a 4--dimensional space which admits a triplet of 
K\"ahler structures $d J = 0 = d K$.
This can be a hyper--K\"ahler space or a Calabi--Yau 2--fold, i.e. 
$K3$.
We then construct a 7--dimensional space by taking the simple product 
with three circles $K_3 \times S^1 \times S^1 \times S^1$. 
The first non--trivial ingredient now added in order to 
introduce some flux is the twist of the metric on 
two of the above circles $K_3 \times S_T^1 \times S_T^1 \times S^1$. 
Explicitly, one makes a non--trivial fibration of two of such circles on 
the base space adding to their standard einbein an extra 1--form valued 
on $K3$. However, 
this manifold does not allow yet for a warp--factor different from zero.

In order to achieve this, one can conformally rescale the 6--dimensional space 
given by the $K3$ and the two circles fibered over it.
The function which is used in such rescaling will give the warp factor 
and should be chosen to depend only on the coordinate of the extra 
$S^1$.
The resulting 7--dimensional metric is
\begin{equation}
ds_7^2 = e^{-2\Delta(y_3)}\left[ds_{K3}^2 + \left(dy_1 + 
\beta_1\right)^2 + \left(dy_2 + \beta_2\right)^2\right] + dy_3^2\,,
\label{eqseven}
\end{equation}
where $y_i$ parameterize the three circles and $\beta_i$ are 1--forms 
valued on $K3$.
This becomes a full solution of M--theory by using the 4--form 
flux
\begin{equation}
\begin{array}{rcl}
G &=& e^{-2\Delta(y_3)} \left[\left(dy_1 + \beta_1\right)\wedge dy_3 \wedge 
\omega_{1} - \left(dy_2 + 
\beta_2\right) \wedge dy_3 \wedge \omega_{2}\right]\\[2mm]
&+& \displaystyle \frac12 \Delta' \, e^{-4\Delta}\left[J\wedge  J + 2 J 
\wedge \left(dy_1 + 
\beta_1\right) \wedge \left(dy_2 + \beta_2\right)\right]\,,
\end{array}
\label{eqGM}
\end{equation}
where $J$ is now the K\"ahler form and 
$\omega_i = d\beta_i$ are harmonic $(1,1)$--forms on $K3$.
This flux is not closed for a generic choice of $\Delta$, but this 
will then be fixed by the specific setup one uses in order to build 
the solution.
In the case of no source contributions to the 4--form Bianchi identity, 
i.e. $dG = 0$, one fixes the $\Delta$ dependence as
\begin{equation}
e^\Delta = c_2 \, \left(c_1 + 4 y_3\right)^{-1/4}\,,
\label{eqedel}
\end{equation}
with $c_1$, $c_2$ real integration constants.

Using one of the twisted circles as 11th coordinate, 
the reduction to type IIA gives a solution with all fluxes   
turned on.
This is actually a forced reduction since translations along $y_3$ 
are not isometries of the metric (\ref{eqseven}).
Solutions of this type arise as T--duals of type IIB compactifications 
with 3--form fluxes on $K3 \times T^2$, as shown in \cite{Tripathy:2002qw}.
We notice however that from \cite{Tripathy:2002qw} it is not clear how 
the terms proportional to $\Delta'$ can be recovered since the warp 
factor is neglected.
It is natural to expect that they arise as contributions from the 
dualization of the 5--form flux of the type IIB solution which depends 
on the derivative of the dilaton.


Another type of solution, based on twisted tori was 
presented in \cite{Kachru:2002sk}.
This solution was obtained as
T--dual of ${\mathbb T}^6/{\mathbb Z}_2$, which is a solution of the type 
IIB theory in the presence of 3--form fluxes.
The resulting manifold is a nilmanifold \cite{salamon} and it should be
a consistent background when all the fluxes of type IIA 
are turned on.
Its metric in the string frame in 10 dimensions 
is given by ${\mathbb T}^3 \times {\mathbb T}^3$, where one 
of the two tori is twisted
\begin{equation} 
d\tilde s_6^2 = e^{-2\phi/3}\,\left(dx_1 + 2 x_2 \, dx_3\right)^2 + 
e^{2\phi/3}\,\left(dx^2_2 + dx_3^2 + 
\sum_{i=1}^3 \, dy_i^2\right)\,.
\label{eqmetrictorus}
\end{equation}
For simplicity we set to 1 the radii of the various tori.
The fluxes  are 
\begin{eqnarray}
F & = & 2 \, dx_2 \wedge dy_3\,, \label{eqFsol}\\[2mm]
H & = & 2 \, dy_1 \wedge dy_2 \wedge dx_3 \,,\label{eqHsol}\\[2mm]
G_{IIA} & =  & 2\, \left(dx_1 + 2 x_2 \, dx_3\right)\wedge  dy_1 \wedge dy_2 \wedge 
dy_3\,. \label{eqGsol}
\end{eqnarray}
The four--flux in 11 dimensions can be reconstructed from 
(\ref{eqFsol})--(\ref{eqGsol}) as 
\begin{equation}
G =  2 \left(dx_1 + 2 x_2 \, dx_3\right)\wedge  dy_1 \wedge dy_2 \wedge 
dy_3 + \tau 
\wedge dy_1 \wedge dy_2 \wedge dx_3 \,,
\label{eqss}
\end{equation}
where $\tau = dz + 2 x_2\, dy_3$ and $z$ is the uplift circle coordinate.
The flux (\ref{eqss}) can also be written as $G = v \wedge U$ and 
therefore the only non--trivial component 
of the flux is given by the $(2,1)+(1,2)$--form $U$ or its dual $S$.
Also in this case we can apply the same comments as above concerning 
the fact that the solution was discussed in \cite{Kachru:2002sk} 
by using the $e^{-2\phi} \sim 1$ approximation.


\bigskip \bigskip

\noindent
{\bf Acknowledgments}

\medskip

We would like to thank G. L. Cardoso, A. Ceresole, S. Chiossi, G.
Curio, D. L\"ust, D. Martelli, G. Papadopoulos, I. Pesando, A.
Tomasiello and D. Waldram for valuable discussions. 
Our work is supported in part by the European Community's Human Potential
Programme under contract HPRN--CT--2000--00131 Quantum Spacetime.
The work of N. P. is supported by the Deutsche Forschungsgemeinschaft under the
project number DFG Lu 419/7-2.



\begin{thebibliography}{100}

\bibitem{Frey:2003tf}
{Frey, Andrew R.}, {\it Warped strings: Self-dual flux and contemporary
  compactifications},  \href{http://arXiv.org/abs/hep-th/0308156}{{\tt
  hep-th/0308156}}.

\bibitem{uno}
P.~Candelas, G.~T. Horowitz, A.~Strominger and E.~Witten, {\it Vacuum
  configurations for superstrings},  {\em Nucl. Phys.} {\bf B258} (1985)
  46--74;
C.~M. Hull, {\it Compactifications of the heterotic superstring},  {\em Phys.
  Lett.} {\bf B178} (1986) 357;
A.~Strominger, {\it Superstrings with torsion},  {\em Nucl. Phys.} {\bf B274}
  (1986) 253;
B.~De~Wit, D.~J. Smit and N.~D. Hari~Dass, {\it Residual supersymmetry of
  compactified d = 10 supergravity},  {\em Nucl. Phys.} {\bf B283} (1987) 165.

\bibitem{Gauntlett:2001ur}
J.~P. Gauntlett, N.-W. Kim, D.~Martelli and D.~Waldram, {\it Fivebranes wrapped
  on {SLAG} three-cycles and related geometry},  {\em JHEP} {\bf 11} (2001) 018
  [\href{http://arXiv.org/abs/hep-th/0110034}{{\tt hep-th/0110034}}].


\bibitem{collect1}
J.~Polchinski and A.~Strominger, {\it New vacua for type {II} string theory},
  {\em Phys. Lett.} {\bf B388} (1996) 736--742
  [\href{http://arXiv.org/abs/http://arXiv.org/abs/hep-th/9510227}{{\tt
  hep-th/9510227}}];
  S.~Gukov, C.~Vafa and E.~Witten,
  {\it CFT's from Calabi-Yau four-folds,}
  Nucl.\ Phys.\ B {\bf 584} (2000) 69
  [Erratum-ibid.\ B {\bf 608} (2001) 477]
  [arXiv:hep-th/9906070];
  S.~Gukov,
  Nucl.\ Phys.\ B {\bf 574} (2000) 169
  [arXiv:hep-th/9911011];
T.~R. Taylor and C.~Vafa, {\it {$RR$} flux on {Calabi-Yau} and partial
  supersymmetry breaking},  {\em Phys. Lett.} {\bf B474} (2000) 130--137
  [\href{http://arXiv.org/abs/http://arXiv.org/abs/hep-th/9912152}{{\tt
  hep-th/9912152}}];
P.~Mayr, {\it On supersymmetry breaking in string theory and its realization in
  brane worlds},  {\em Nucl. Phys.} {\bf B593} (2001) 99--126
  [\href{http://arXiv.org/abs/http://arXiv.org/abs/hep-th/0003198}{{\tt
  hep-th/0003198}}];
G.~Curio, A.~Klemm, D.~L{\"u}st and S.~Theisen, {\it On the vacuum structure of
  type {II} string compactifications on {Calabi-Yau} spaces with
  {$H$}--fluxes},  {\em Nucl. Phys.} {\bf B609} (2001) 3--45
  [\href{http://arXiv.org/abs/http://arXiv.org/abs/hep-th/0012213}{{\tt
  hep-th/0012213}}];
S.~B. Giddings, S.~Kachru and J.~Polchinski, {\it Hierarchies from fluxes in
  string compactifications},  {\em Phys. Rev.} {\bf D66} (2002) 106006
  [\href{http://arXiv.org/abs/hep-th/0105097}{{\tt hep-th/0105097}}];
G.~Curio, A.~Klemm, B.~K{\"o}rs and D.~L{\"u}st, {\it Fluxes in heterotic and
  type {II} string compactifications},  {\em Nucl. Phys.} {\bf B620} (2002)
  237--258
  [\href{http://arXiv.org/abs/http://arXiv.org/abs/hep-th/0106155}{{\tt
  hep-th/0106155}}];
G.~Dall'Agata, {\it Type {IIB} supergravity compactified on a {Calabi-Yau}
  manifold with {H-fluxes}},  {\em JHEP} {\bf 11} (2001) 005
  [\href{http://arXiv.org/abs/hep-th/0107264}{{\tt hep-th/0107264}}];
G.~Curio, B.~K{\"o}rs and D.~L{\"u}st, {\it Fluxes and branes in type {II}
  vacua and {M}-theory geometry with {$G_2$} and {$Spin(7)$} holonomy},  {\em
  Nucl. Phys.} {\bf B636} (2002) 197--224
  [\href{http://arXiv.org/abs/http://arXiv.org/abs/hep-th/0111165}{{\tt
  hep-th/0111165}}];
S.~Kachru, M.~B. Schulz and S.~Trivedi, {\it Moduli stabilization from fluxes
  in a simple {IIB} orientifold},
  \href{http://arXiv.org/abs/http://arXiv.org/abs/hep-th/0201028}{{\tt
  hep-th/0201028}};
 J.~Louis and A.~Micu, {\it Type {II} theories compactified on {Calabi-Yau}
    threefolds in the presence of background fluxes},  {\em Nucl. Phys.} {\bf
    B635} (2002) 395--431
    [\href{http://arXiv.org/abs/http://arXiv.org/abs/hep-th/0202168}{{\tt
    hep-th/0202168}}];
J.~P. Gauntlett, D.~Martelli, S.~Pakis and D.~Waldram, {\it {G-structures} and
  wrapped {NS5-branes}},  \href{http://arXiv.org/abs/hep-th/0205050}{{\tt
  hep-th/0205050}};
S.~Gurrieri, J.~Louis, A.~Micu and D.~Waldram, {\it Mirror symmetry in
  generalized {Calabi-Yau} compactifications},  {\em Nucl. Phys.} {\bf B654}
  (2003) 61--113 [\href{http://arXiv.org/abs/hep-th/0211102}{{\tt
  hep-th/0211102}}];
  M.~Berg, M.~Haack and B.~Kors, {\it An orientifold with fluxes and branes via
    {T-duality}},  {\em Nucl. Phys.} {\bf B669} (2003) 3--56
    [\href{http://arXiv.org/abs/hep-th/0305183}{{\tt hep-th/0305183}}].

\bibitem{Kachru:2002sk}
S.~Kachru, M.~B. Schulz, P.~K. Tripathy and S.~P. Trivedi, {\it New
  supersymmetric string compactifications},  {\em JHEP} {\bf 03} (2003) 061
  [\href{http://arXiv.org/abs/hep-th/0211182}{{\tt hep-th/0211182}}].

\bibitem{Tripathy:2002qw}   
P.~K. Tripathy and S.~P. Trivedi, {\it Compactification with flux on {K3} and
    tori},  \href{http://arXiv.org/abs/hep-th/0301139}{{\tt hep-th/0301139}}.


\bibitem{collect2}
J.~Louis and A.~Micu, {\it Heterotic string theory with background fluxes},
  {\em Nucl. Phys.} {\bf B626} (2002) 26--52
  [\href{http://arXiv.org/abs/http://arXiv.org/abs/hep-th/0110187}{{\tt
  hep-th/0110187}}];
K.~Becker and K.~Dasgupta, {\it Heterotic strings with torsion},
  \href{http://arXiv.org/abs/http://arXiv.org/abs/hep-th/0209077}{{\tt
  hep-th/0209077}};
  K.~Becker, M.~Becker, K.~Dasgupta and S.~Prokushkin, {\it Properties of
    heterotic vacua from superpotentials},  {\em Nucl. Phys.} {\bf B666} (2003)
    144--174 [\href{http://arXiv.org/abs/hep-th/0304001}{{\tt hep-th/0304001}}];
  J.~Gillard, G.~Papadopoulos and D.~Tsimpis, {\it Anomaly, fluxes and (2,0)
    heterotic-string compactifications},  {\em JHEP} {\bf 06} (2003) 035
    [\href{http://arXiv.org/abs/hep-th/0304126}{{\tt hep-th/0304126}}];
  G.~L. Cardoso, G.~Curio, G.~Dall'Agata and D.~L\"ust, {\it {BPS} action and
    superpotential for heterotic string compactifications with fluxes},
    \href{http://arXiv.org/abs/hep-th/0306088}{{\tt hep-th/0306088}};
  G.~L. Cardoso, G.~Curio, G.~Dall'Agata and D.~L\"ust, {\it Heterotic string
    theory on {non-K\"ahler} manifolds with {H}- flux and gaugino condensate},
    \href{http://arXiv.org/abs/hep-th/0310021}{{\tt hep-th/0310021}};
    K.~Becker, M.~Becker, K.~Dasgupta and P.~S. Green, {\it Compactifications of
      heterotic theory on {non-K\"ahler} complex manifolds. {I}},
      \href{http://arXiv.org/abs/hep-th/0301161}{{\tt hep-th/0301161}};
  K.~Becker, M.~Becker, K.~Dasgupta, P.~S. Green and E.~Sharpe, {\it
    Compactifications of heterotic strings on {non-K\"ahler} complex manifolds.
    {II}},  \href{http://arXiv.org/abs/hep-th/0310058}{{\tt hep-th/0310058}};
  S.~Gukov, S.~Kachru, X.~Liu and L.~McAllister, {\it Heterotic moduli
    stabilization with fractional {Chern-Simons} invariants},
    \href{http://arXiv.org/abs/hep-th/0310159}{{\tt hep-th/0310159}};
    M.~Serone and M.~Trapletti,
    {\it String vacua with flux from freely-acting obifolds,}
    {{\tt hep-th/0310245}}.
  
 
\bibitem{Cardoso:2002hd}
G.~L. Cardoso, G.~Curio, G.~Dall'Agata, D.~L\"ust, P.~Manousselis and
  G.~Zoupanos, {\it {Non-K\"ahler} string backgrounds and their five torsion
  classes},  {\em Nucl. Phys.} {\bf B652} (2003) 5--34
  [\href{http://arXiv.org/abs/hep-th/0211118}{{\tt hep-th/0211118}}].



\bibitem{Gauntlett:2003cy}
J.~P. Gauntlett, D.~Martelli and D.~Waldram, {\it Superstrings with intrinsic
  torsion},  \href{http://arXiv.org/abs/hep-th/0302158}{{\tt hep-th/0302158}}.



\bibitem{Ibanez:2001dj}
L.~E. Ibanez, {\it Standard model engineering with intersecting branes},
  \href{http://arXiv.org/abs/hep-ph/0109082}{{\tt hep-ph/0109082}}.

\bibitem{Blumenhagen:2002vp}
R.~Blumenhagen, V.~Braun, B.~Kors and D.~L\"ust, {\it The standard model on the
  quintic},  \href{http://arXiv.org/abs/hep-th/0210083}{{\tt hep-th/0210083}}.

\bibitem{Cvetic}
  M.~Cvetic, G.~Shiu and A.~M.~Uranga,
  {\it Chiral four-dimensional N = 1 supersymmetric type IIA orientifolds from
  intersecting D6-branes,}
  Nucl.\ Phys. {\bf B615}, 3 (2001)
  {{\tt [hep-th/0107166].}}


\bibitem{Gauntlett:2002fz}
J.~P. Gauntlett and S.~Pakis, {\it The geometry of {D} = 11 {Killing} spinors},
   \href{http://arXiv.org/abs/hep-th/0212008}{{\tt hep-th/0212008}}.

\bibitem{Gauntlett:2003new}
J.~Gauntlett, J.~Gutowski and S.~Pakis, {\it The geometry of {D=11} null
  killing spinors},  \href{http://arXiv.org/abs/hep-th/0311112}{{\tt
  hep-th/0311112}}.

\bibitem{Behrndt:2003uq}
K.~Behrndt and C.~Jeschek, {\it Fluxes in {M}-theory on 7-manifolds and {G}
  structures},  {\em JHEP} {\bf 04} (2003) 002
  [\href{http://arXiv.org/abs/hep-th/0302047}{{\tt hep-th/0302047}}].

\bibitem{Kaste:2003zd}
P.~Kaste, R.~Minasian and A.~Tomasiello, {\it Supersymmetric {M-theory}
  compactifications with fluxes on seven-manifolds and {G-structures}},
  \href{http://arXiv.org/abs/hep-th/0303127}{{\tt hep-th/0303127}}.

\bibitem{collect3}
K.~Becker and M.~Becker, {\it {M-Theory} on eight-manifolds},  {\em Nucl.
  Phys.} {\bf B477} (1996) 155--167
  [\href{http://arXiv.org/abs/hep-th/9605053}{{\tt hep-th/9605053}}];
K.~Dasgupta, G.~Rajesh and S.~Sethi, {\it {M} theory, orientifolds and
  {G-flux}},  {\em JHEP} {\bf 08} (1999) 023
  [\href{http://arXiv.org/abs/hep-th/9908088}{{\tt hep-th/9908088}}];
K.~Becker, {\it A note on compactifications on {Spin(7)-holonomy} manifolds},
  {\em JHEP} {\bf 05} (2001) 003
  [\href{http://arXiv.org/abs/hep-th/0011114}{{\tt hep-th/0011114}}];
B.~Acharya, X.~De~La~Ossa and S.~Gukov, {\it {G}-flux, supersymmetry and
  {Spin(7)} manifolds},  {\em JHEP} {\bf 09} (2002) 047
  [\href{http://arXiv.org/abs/hep-th/0201227}{{\tt hep-th/0201227}}];
B.~S. Acharya, {\it Compactification with flux and {Yukawa} hierarchies},
  \href{http://arXiv.org/abs/hep-th/0303234}{{\tt hep-th/0303234}};
D.~Martelli and J.~Sparks, {\it {G-structures}, fluxes and calibrations in
  {M-theory}},  {\em Phys. Rev.} {\bf D68} (2003) 085014
  [\href{http://arXiv.org/abs/hep-th/0306225}{{\tt hep-th/0306225}}].

\bibitem{Kaste:2002xs}
P.~Kaste, R.~Minasian, M.~Petrini and A.~Tomasiello, {\it {Kaluza-Klein}
  bundles and manifolds of exceptional holonomy},  {\em JHEP} {\bf 09} (2002)
  033 [\href{http://arXiv.org/abs/hep-th/0206213}{{\tt hep-th/0206213}}].

\bibitem{Kaste:2003dh}
P.~Kaste, R.~Minasian, M.~Petrini and A.~Tomasiello, {\it Nontrivial {RR}
  two-form field strength and {SU(3)}-structure},  {\em Fortsch. Phys.} {\bf
  51} (2003) 764--768 [\href{http://arXiv.org/abs/hep-th/0301063}{{\tt
  hep-th/0301063}}].

\bibitem{Behrndt:2003ih}
K.~Behrndt and M.~Cvetic, {\it Supersymmetric intersecting d6-branes and fluxes
  in massive type iia string theory},
  \href{http://arXiv.org/abs/hep-th/0308045}{{\tt hep-th/0308045}}.

\bibitem{Friedrich:1995Dp}
T.~Friedrich, I.~Kath, A.~Moroianu and U.~Semmelmann, {\it On nearly parallel
  {$G_2$} structures}, . SFB-288-162.

\bibitem{Hitchin}
N.~Hitchin, {\it Stable forms and special metrics},  in {\em Global
  Differential Geometry: The mathematical legacy of Alfred Gray}, pp.~70--89,
  Ams, 2001.
\newblock \href{http://arXiv.org/abs/math-dg/0107101}{{\tt math-dg/0107101}}.

\bibitem{Fayyazuddin:1999zu}
A.~Fayyazuddin and D.~J. Smith, {\it Localized intersections of {M5-branes} and
  four-dimensional superconformal field theories},  {\em JHEP} {\bf 04} (1999)
  030 [\href{http://arXiv.org/abs/hep-th/9902210}{{\tt hep-th/9902210}}].

\bibitem{Brinne:2000nf}
B.~Brinne, A.~Fayyazuddin, T.~Z. Husain and D.~J. Smith, {\it {N = 1 M5-brane}
  geometries},  {\em JHEP} {\bf 03} (2001) 052
  [\href{http://arXiv.org/abs/hep-th/0012194}{{\tt hep-th/0012194}}].

\bibitem{Smith:2002wn}
D.~J. Smith, {\it Intersecting brane solutions in string and {M-theory}},  {\em
  Class. Quant. Grav.} {\bf 20} (2003) R233
  [\href{http://arXiv.org/abs/hep-th/0210157}{{\tt hep-th/0210157}}].

\bibitem{Husain:2003df}  T.~Z.~Husain,
  {\it That's a wrap!,}  JHEP {\bf 0304} (2003) 053  [arXiv:hep-th/0302071].

\bibitem{ClausKlaus}
K.~Behrndt and C.~Jeschek, {\it Fluxes in {M-theory} on 7-manifolds:
  {$G$-}structures and superpotential},
  \href{http://arXiv.org/abs/hep-th/0311119}{{\tt hep-th/0311119}}.

\bibitem{last}
S. Fidanza, R. Minasian and A. Tomasiello, {\it Mirror symmetric SU(3)-structure manifolds with NS fluxes
}, {{\tt hep-th/0311122}}.


\bibitem{Joyce}
D.~Joyce, {\em Compact Manifolds with Special Holonomy}.
\newblock Oxford University Press, oxford mathematical monographs~ed., 2000.

\bibitem{chiossi}
S.~Chiossi and S.~Salamon, {\it The intrinsic torsion of {$SU(3)$} and {$G_2$}
  structures},  in {\em Differential Geometry, Valencia 2001}, (River Edge,
  {NJ}), pp.~115--133, World Sci. Publishing, 2002.
\newblock \href{http://arXiv.org/abs/math.DG/0202282}{{\tt math.DG/0202282}}.

\bibitem{Becker:2000rz}
K.~Becker and M.~Becker, {\it Compactifying {M}-theory to four dimensions},
  {\em JHEP} {\bf 11} (2000) 029
  [\href{http://arXiv.org/abs/hep-th/0010282}{{\tt hep-th/0010282}}].

\bibitem{Curio:2003ur} 
G.~Curio and A.~Krause,
{\it Four-flux and warped heterotic M-theory compactifications,}
Nucl.\ Phys.\ B {\bf 602} (2001) 172
{\tt [hep-th/0012152]};
G.~Curio and A.~Krause,
{\it Enlarging the parameter space of heterotic M-theory flux compactifications to phenomenological viability,}
  {\tt hep-th/0308202}.

\bibitem{Gutowski:1999iu}
J.~Gutowski and G.~Papadopoulos, {\it {AdS} calibrations},  {\em Phys. Lett.}
  {\bf B462} (1999) 81--88 [\href{http://arXiv.org/abs/hep-th/9902034}{{\tt
  hep-th/9902034}}].

\bibitem{Gutowski:1999tu}
J.~Gutowski, G.~Papadopoulos and P.~K. Townsend, {\it Supersymmetry and
  generalized calibrations},  {\em Phys. Rev.} {\bf D60} (1999) 106006
  [\href{http://arXiv.org/abs/hep-th/9905156}{{\tt hep-th/9905156}}].

\bibitem{Barwald:1999ux}
O.~Barwald, N.~D. Lambert and P.~C. West, {\it A calibration bound for the
  {M}-theory fivebrane},  {\em Phys. Lett.} {\bf B463} (1999) 33--40
  [\href{http://arXiv.org/abs/hep-th/9907170}{{\tt hep-th/9907170}}].

\bibitem{Gutowski:2002bc}
J.~Gutowski, S.~Ivanov and G.~Papadopoulos, {\it Deformations of generalized
  calibrations and compact non--{K\"ahler} manifolds with vanishing first
  {Chern} class},
  \href{http://arXiv.org/abs/http://arXiv.org/abs/math.dg/0205012}{{\tt
  math.dg/0205012}}.

\bibitem{AGS}
E.~Abbena, S.~Garbiero and S.~Salamon, {\it Almost hermitian geometry on six
  dimensional nilmanifolds},  {\em Ann. Sc. Norm. Sup.} (2000), no.~30 147--170
  [\href{http://arXiv.org/abs/math.DG/0007066}{{\tt math.DG/0007066}}].

\bibitem{salamon}
S.~Salamon, {\it Complex structures on nilpotent {Lie} algebras},  {\em J Pure
  Appl Algebra} (1998) [\href{http://arXiv.org/abs/math.DG/9808025}{{\tt
  math.DG/9808025}}].

\bibitem{Goldstein:2002pg}
E.~Goldstein and S.~Prokushkin, {\it Geometric model for complex {non-K\"ahler}
  manifolds with {SU(3)} structure},
  \href{http://arXiv.org/abs/hep-th/0212307}{{\tt hep-th/0212307}}.

\end{thebibliography}

\providecommand{\href}[2]{#2}\begingroup
\endgroup

\end{document}